\documentclass{LMCS}

\def\dOi{10(3:22)2014}
\lmcsheading%
{\dOi}
{1--28}
{}
{}
{Oct.~27, 2010}
{Sep.~18, 2014}
{}

\ACMCCS{[{\bf Theory of computation}]:Formal languages and automata
  theory---Grammars and context-free languages; Formal languages and automata
  theory---Formalisms---Rewrite systems; Logic}


\usepackage{hyperref}
\usepackage{tikz}
\usepackage{amssymb}

\newcommand{\NN}{\mathbb{N}}
\newcommand{\QQ}{\mathbb{Q}}
\newcommand{\RR}{\mathbb{R}}
\newcommand{\ZZ}{\mathbb{Z}}
\newcommand{\m}[1]{\mathsf{#1}}
\newcommand{\seq}[2][n]{{#2_1},\dots,{#2_{#1}}}
\renewcommand{\geq}{\geqslant}

\begin{document}

\title[Polynomial Interpretations Revisited]
{Polynomial Interpretations over the Natural, Rational and Real Numbers
Revisited}

\author[F.~Neurauter]{Friedrich Neurauter\rsuper a}
\address{{\lsuper a}TINETZ-Stromnetz Tirol AG}
\email{friedrich.neurauter@aon.at}

\author[A.~Middeldorp]{Aart Middeldorp\rsuper b}
\address{{\lsuper b}Institute of Computer Science \newline
University of Innsbruck, Austria}
\email{aart.middeldorp@uibk.ac.at}


\begin{abstract}
Polynomial interpretations are a useful technique for proving
termination of term rewrite systems. They come in various flavors:
polynomial interpretations with real, rational and integer coefficients.
As to their relationship with respect to termination proving power,
Lucas managed to prove in 2006 that there are rewrite systems that can
be shown polynomially terminating by polynomial interpretations with real
(algebraic) coefficients, but cannot be shown polynomially terminating
using polynomials with rational coefficients only.
He also proved the corresponding statement regarding the use of rational
coefficients versus integer coefficients.
In this article we extend these results, thereby giving the full picture
of the relationship between the aforementioned variants of polynomial
interpretations. In particular, we show that polynomial
interpretations with real or rational coefficients do not subsume
polynomial interpretations with integer coefficients.
Our results hold also for incremental termination proofs with
polynomial interpretations.
\end{abstract}

\maketitle

\section{Introduction}
\label{sect:intro}

Polynomial interpretations are a simple yet useful technique for proving
termination of term rewrite systems (TRSs, for short). While originally
conceived in the late seventies by
Lankford \cite{L79} as a means for establishing direct termination
proofs,
polynomial interpretations are nowadays often used in the context of the
dependency pair (DP) framework \cite{AG00,GTSF06,HM07}.
In the classical approach of Lankford, one considers polynomials with
integer coefficients inducing polynomial algebras over the well-founded
domain of the natural numbers.
To be precise, every $n$-ary function symbol $f$ is interpreted by a
polynomial $P_f$ in $n$ indeterminates with
integer coefficients, which induces a mapping or \emph{interpretation}
from terms to integer numbers in the obvious way. In order to conclude
termination of a given TRS, three conditions have to be satisfied.
First, every polynomial must be \emph{well-defined}, i.e., it must induce
a well-defined polynomial function
$f_\NN\colon \NN^n \to \NN$ over the natural numbers. In addition, the
interpretation functions $f_\NN$ are required to be
\emph{strictly monotone} in all arguments. Finally, one has to show
\emph{compatibility} of the interpretation with the given TRS. More
precisely,
for every rewrite rule $\ell \to r$, the polynomial $P_\ell$ associated
with the left-hand side must be greater than $P_r$, the corresponding
polynomial of the right-hand side, i.e., $P_\ell > P_r$ for all values
of the indeterminates.

Already back in the seventies, an alternative approach using polynomials
with real coefficients instead of integers was proposed by
Dershowitz~\cite{D79}. However, as the real numbers $\RR$ equipped
with the standard order $>_\RR$ are not well-founded,
a subterm property is explicitly required to ensure well-foundedness.
It was not until 2005 that this limitation was overcome, when
Lucas \cite{L05} presented a framework for proving polynomial termination
over the real numbers, where well-foundedness is basically
achieved by replacing $>_\RR$ with a new ordering $>_{\RR,\delta}$
requiring comparisons between terms to not be below a given positive real
number $\delta$. Moreover, this framework also facilitates polynomial
interpretations over the rational numbers.

Thus, one can distinguish three variants of polynomial interpretations,
polynomial interpretations with real, rational and integer coefficients,
and the obvious question is:
what is their relationship with regard to termination proving power?
For Knuth-Bendix orders it is known~\cite{KV03,L01} that extending the
range of the underlying weight function from natural numbers to
non-negative reals does not result in an increase in termination proving
power. In 2006 Lucas~\cite{L06} proved that there are TRSs that can be
shown polynomially terminating by polynomial interpretations with
rational coefficients, but cannot be shown polynomially terminating using
polynomials with integer coefficients only. Likewise, he proved that
there are TRSs that can be handled by polynomial interpretations with
real (algebraic)
coefficients, but cannot be handled by polynomial interpretations
with rational coefficients.

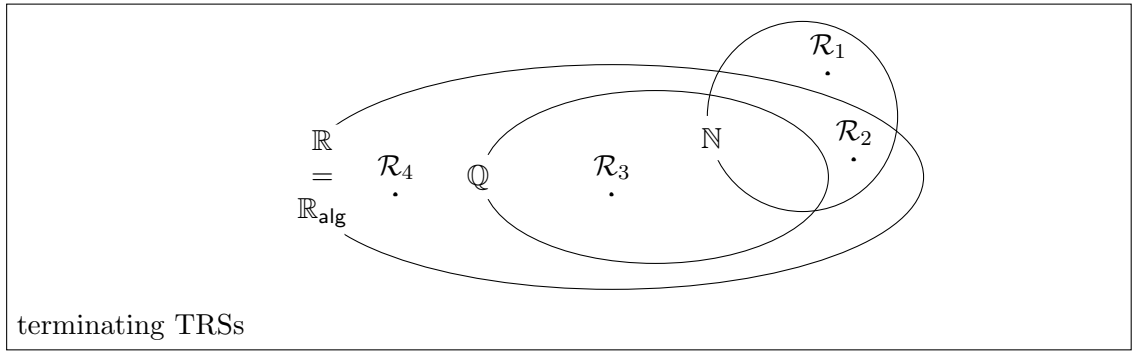
\begin{figure}[bt]
\begin{center}
\begin{tikzpicture}[scale=1.15]
\draw[fill=white] (0,0) node[anchor=south west] {terminating TRSs}
rectangle (13cm,4cm);
\begin{scope}
\path[clip] (7cm,20mm) ellipse (36mm and 13mm);
\path[fill=white] (86mm,25mm) ellipse (9mm and 9mm);
\end{scope}
\draw[fill=white] (75mm,20mm) ellipse (2cm and 10mm);
\draw (7cm,20mm) ellipse (36mm and 13mm);
\draw[fill=white,draw=white] (32mm,12mm) rectangle ++(7mm,14mm);
\draw (36.5mm,20mm)
  node {$\begin{array}{c} \RR \\ = \\ \RR_{\m{alg}} \end{array}$};
\draw[fill=white,draw=white] (52mm,17.5mm) rectangle ++(5mm,5mm);
\draw (54.5mm,20mm) node {$\QQ$};
\draw (92mm,27mm) ellipse (11mm and 11mm);
\draw[fill=white,draw=white] (79mm,22mm) rectangle ++(5mm,5mm);
\draw (81.5mm,24.5mm) node {$\NN$};
\draw (45mm,20mm)
  node {$\stackrel{\raisebox{1mm}{$\mathcal{R}_4$}}{\pmb{\cdot}}$};
\draw (70mm,20mm)
  node {$\stackrel{\raisebox{1mm}{$\mathcal{R}_3$}}{\pmb{\cdot}}$};
\draw (98mm,24mm)
  node {$\stackrel{\raisebox{1mm}{$\mathcal{R}_2$}}{\pmb{\cdot}}$};
\draw (95mm,34mm)
  node {$\stackrel{\raisebox{1mm}{$\mathcal{R}_1$}}{\pmb{\cdot}}$};
\end{tikzpicture}
\end{center}
\caption{Comparison.}
\label{fig:summary}
\end{figure}

In this article we extend these results and give a complete comparison
between the various notions of polynomial termination.\footnote{%
Readers familiar with Lucas~\cite{L06} should note that we use a
different definition of polynomial termination over the reals and
rationals, cf.\ Remark~\ref{Lucas}.}
In general, the situation turns out to be
as depicted in Figure~\ref{fig:summary}, which illustrates both
our results and the earlier results of Lucas~\cite{L06}.
In particular, we prove that polynomial interpretations with real
coefficients
subsume polynomial interpretations with rational coefficients.
Moreover, we show that polynomial interpretations with real or rational
coefficients do not subsume polynomial interpretations with integer
coefficients by exhibiting the TRS $\mathcal{R}_1$ in
Section~\ref{sect:nvsr}.
Likewise, we prove that there are TRSs
that can be shown terminating by polynomial interpretations with real
coefficients as well as by polynomial interpretations with integer
coefficients, but cannot be shown terminating using polynomials with
rational coefficients only, by exhibiting the TRS $\mathcal{R}_2$ in
Section~\ref{sect:nrvsq}. The TRSs $\mathcal{R}_3$ and
$\mathcal{R}_4$ can be found in Section~\ref{sect:incrementality}.

The remainder of this article is organized as follows. In
Section~\ref{sect:prelim}, we introduce some preliminary definitions
and terminology concerning polynomials and polynomial interpretations.
In Section~\ref{sect:qvsr}, we show that polynomial
interpretations with real coefficients subsume polynomial interpretations
with rational coefficients. We further show that for polynomial
interpretations over the reals, it suffices to consider real algebraic
numbers as interpretation domain.
Section~\ref{sect:nvsr} is dedicated to showing that polynomial
interpretations with real or rational coefficients do not subsume
polynomial interpretations with integer coefficients.
Then, in Section~\ref{sect:nrvsq}, we present a TRS
that can be handled by a polynomial interpretation with real coefficients
as well as by a polynomial interpretation with integer coefficients, but
cannot be handled using polynomials with rational coefficients.
In Section~\ref{sect:incrementality}, we show that the relationships
in Figure~\ref{fig:summary} remain true if incremental termination
proofs with polynomial interpretations are considered.
We conclude in Section~\ref{sect:conclusion}.

This paper is an extended version of \cite{FNAM-RTA10}, which
contained the result of Section~\ref{sect:nvsr}.
The results in Sections~\ref{sect:qvsr}, \ref{sect:nrvsq}
and~\ref{sect:incrementality} are new.

\section{Preliminaries}
\label{sect:prelim}

As usual, we denote by $\NN$, $\ZZ$, $\QQ$ and $\RR$ the sets of natural,
integer, rational and real numbers, respectively. An \emph{irrational}
number is a real number, which is not in $\QQ$. Given some
$D \in \{ \NN, \ZZ, \QQ, \RR \}$ and $m \in D$,
$>_D$ denotes the standard order of the respective domain and
$D_m := \{ x \in D \mid x \geq m \}$.
A sequence of real numbers $(x_n)_{n \in \NN}$ \emph{converges} to the
\emph{limit} $x$ if for every real number $\varepsilon > 0$ there exists
a natural number $N$ such that the absolute distance $|x_n - x|$ is less
than $\varepsilon$ for all $n > N$; we denote this by
$\lim_{n \to \infty} x_n = x$. As convergence
in $\RR^k$ is equivalent to componentwise convergence, we use the same
notation also for limits of converging sequences of vectors of real
numbers $(\vec{x}_n \in \RR^k)_{n \in \NN}$.
A real function $f\colon \RR^k \to \RR$ is \emph{continuous} in $\RR^k$
if for every converging sequence $(\vec{x}_n \in \RR^k)_{n \in \NN}$
it holds that
$\lim_{n \to \infty} f(\vec{x}_n) = f(\lim_{n \to \infty} \vec{x}_n)$.
Finally, as $\QQ$ is dense in $\RR$, every real number is a
rational number or the limit of a converging sequence of rational
numbers.

\subsection*{Polynomials}

For any ring $R$ (e.g. $\ZZ$, $\QQ$, $\RR$), we denote the associated
\emph{polynomial ring} in $n$ \emph{indeterminates} $\seq{x}$ by
$R[\seq{x}]$, the elements of which are finite sums of products
of the form $c\cdot x_1^{i_1}x_2^{i_2}\cdots x_n^{i_n}$, where the
\emph{coefficient} $c$ is an element of $R$ and the exponents $\seq{i}$
in the \emph{monomial} $x_1^{i_1}x_2^{i_2}\cdots x_n^{i_n}$
are non-negative integers.
If $c \neq 0$, we call a product
$c\cdot x_1^{i_1}x_2^{i_2}\cdots x_n^{i_n}$ a \emph{term}.
The \emph{degree} of a monomial
is just the sum of its exponents, and the degree of a term is
the degree of its monomial.
An element $P \in R[\seq{x}]$ is called an
\emph{($n$-variate) polynomial with coefficients in $R$}.
For example, the polynomial $2x^2-x+1$ is an element of $\ZZ[x]$,
the ring of all univariate polynomials with integer coefficients.

In the special case $n = 1$, a polynomial $P \in R[x]$ can be written as
follows: $P(x) = \sum_{k=0}^d {a_k x^k}$ ($d \geq 0$).
For the largest $k$ such that $a_k \neq 0$, we call $a_k x^k$ the
\emph{leading term} of $P$, $a_k$ its \emph{leading coefficient}
and $k$ its \emph{degree}, which we denote by $\deg(P) = k$.
A polynomial $P \in R[x]$ is said to be \emph{linear} if $\deg(P) = 1$,
and \emph{quadratic} if $\deg(P) = 2$.

\subsection*{Polynomial Interpretations}

We assume familiarity with the basics of term rewriting and polynomial
interpretations (e.g.\ \cite{BN98,TeReSe}).
The key concept for establishing (direct) termination of TRSs
via polynomial interpretations is the notion of well-founded
monotone algebras as they induce reduction orders on terms.

\begin{defi}
\label{def:ma}
Let $\mathcal{F}$ be a signature, i.e., a set of function symbols equipped
with fixed arities.
An $\mathcal{F}$-\emph{algebra} $\mathcal{A}$ consists if a non-empty
\emph{carrier} set $A$ and a collection of
interpretation functions $f_A\colon A^n \to A$ for
each $n$-ary function symbol $f \in \mathcal{F}$.
The \emph{evaluation} or \emph{interpretation} $[\alpha]_\mathcal{A}(t)$
of a term $t \in \mathcal{T}(\mathcal{F},\mathcal{V})$ with respect to a
variable
assignment $\alpha\colon \mathcal{V} \to A$ is inductively defined as
follows:
\[
[\alpha]_\mathcal{A}(t) =
\begin{cases}
\alpha(t) & \text{if $t \in \mathcal{F}$} \\
f_A([\alpha]_\mathcal{A}(t_1),\dots,[\alpha]_\mathcal{A}(t_n)) &
\text{if $t = f(\seq{t})$}
\end{cases}
\]
Let $\sqsupset$ be a binary relation on $A$.
For $i \in \{ 1, \dots, n \}$, an interpretation function
$f_A \colon A^n \to A$ is \emph{monotone in its $i$-th argument}
with respect to $\sqsupset$ if $a_i \sqsupset b$ implies
\[
f_A(a_1,\dots,a_i,\dots,a_n) \sqsupset f_A(a_1,\dots,b,\dots,a_n)
\]
for all $\seq{a}, b \in A$. It is said to be \emph{monotone} with respect
to $\sqsupset$ if it is monotone in all its arguments.
We define $s \sqsupset_A t$ as
$[\alpha]_\mathcal{A}(s) \sqsupset [\alpha]_\mathcal{A}(t)$
for all assignments $\alpha$.
\end{defi}

In order to pave the way for incremental polynomial termination in
Section~\ref{sect:incrementality}, the following definition is more
general than what is needed for direct termination proofs.

\begin{defi}
\label{def:wma}
Let $(\mathcal{A}, >, \geqslant)$ be an $\mathcal{F}$-algebra together
with two binary relations $>$ and $\geqslant$ on $A$. We say that
$(\mathcal{A},>,\geqslant)$ and a TRS $\mathcal{R}$ are
\emph{(weakly) compatible} if
$\ell >_\mathcal{A} r$ ($\ell \geqslant_\mathcal{A} r$) for each rewrite
rule $\ell \to r \in \mathcal{R}$.
An interpretation function $f_A$ is called strictly (weakly)
monotone if it is monotone with respect to $>$ ($\geqslant$).
The triple $(\mathcal{A}, >, \geqslant)$ (or just $\mathcal{A}$ if $>$
and $\geqslant$ are clear from the context) is a
\emph{weakly (strictly) monotone $\mathcal{F}$-algebra}
if $>$ is well-founded,
${> \cdot \geqslant} \subseteq {>}$ and for each $f \in \mathcal{F}$,
$f_A$ is \emph{weakly (strictly) monotone}.
It is said to be an
\emph{extended monotone $\mathcal{F}$-algebra} if it is both weakly
monotone and strictly monotone.
Finally, we call $(\mathcal{A}, >, \geqslant)$ a
\emph{well-founded monotone $\mathcal{A}$-algebra} if $>$ is a
well-founded order on $A$, $\geqslant$ is its
reflexive closure, and each interpretation function
is strictly monotone.
\end{defi}

It is well-known that well-founded monotone algebras provide a complete
characterization of termination.

\begin{thm}
A TRS is terminating if and only if it is compatible with a well-founded
monotone algebra.
\qed
\end{thm}

\begin{defi}
\label{def:PolyInt_N}
A \emph{polynomial interpretation over $\NN$} for a signature
$\mathcal{F}$ consists of a polynomial $f_\NN \in \ZZ[\seq{x}]$ for every
$n$-ary function symbol $f \in \mathcal{F}$ such that for all
$f \in \mathcal{F}$ the following two properties are satisfied:
\begin{enumerate}
\item
\emph{well-definedness}:
$f_\NN(\seq{x}) \in \NN$ for all $\seq{x} \in \NN$,
\item
\emph{strict monotonicity} of $f_\NN$ in all arguments with respect
to $>_\NN$, the standard order on $\NN$.
\end{enumerate}
Due to well-definedness, each of the polynomials $f_\NN$ induces
a function from $\NN^n$ to $\NN$. Hence, the pair
$\mathcal{N} = (\NN, \{ f_\NN \}_{f \in \mathcal{F}})$ constitutes an
$\mathcal{F}$-algebra over the carrier $\NN$.
Now $(\mathcal{N}, >_\NN, \geqslant_\NN)$
where $\geqslant_\NN$ is the reflexive closure of $>_\NN$ constitutes a
well-founded monotone algebra, and we say that a polynomial
interpretation over $\NN$ is \emph{compatible} with a TRS
$\mathcal{R}$ if the well-founded monotone algebra
$(\mathcal{N}, >_\NN, \geqslant_\NN)$
is compatible with $\mathcal{R}$. Finally, a TRS is
\emph{polynomially terminating over $\NN$} if it admits a
compatible polynomial interpretation over $\NN$.
\end{defi}

In the sequel, we often identify a polynomial
interpretation with its associated $\mathcal{F}$-algebra.

\begin{rem}
\label{rem:isom}
In principle, one could take any set $\NN_m$ (or even $\ZZ_m$) instead of
$\NN$ as the carrier for polynomial interpretations. However, it is
well-known~\cite{TeReSe,CMTU05} that all these sets are
order-isomorphic to $\NN$ and hence do not change the class
of polynomially terminating TRSs.
In other words, a TRS $\mathcal{R}$ is polynomially terminating over $\NN$
if and only if it is polynomially terminating over $\NN_m$.
Thus, we can restrict to $\NN$ as carrier without loss of generality.
\end{rem}

The following simple criterion for strict monotonicity of a
univariate quadratic polynomial will be used in
Sections~\ref{sect:nvsr} and \ref{sect:nrvsq}.

\begin{lem}
\label{lem:nquadmon}
The quadratic polynomial $f_\NN(x) = ax^2 + bx + c$
with $a, b, c \in \ZZ$ is strictly monotone and well-defined if and only
if $a > 0$, $c \geqslant 0$, and $a + b > 0$.
\qed
\end{lem}

Now if one wants to extend the notion of polynomial interpretations to
the rational or real numbers, the main problem one is confronted with
is the non-well-foundedness of these domains with respect to the standard
orders $>_\QQ$ and $>_\RR$. In \cite{H01,L05}, this problem is
overcome by replacing these orders with new non-total orders
$>_{\RR,\delta}$ and
$>_{\QQ,\delta}$, the first of which is defined as
follows: given some fixed positive real number $\delta$,
\[
x >_{\RR,\delta} y \quad:\iff\quad
x-y \geq_\RR \delta \quad \text{for all $x, y \in \RR$.}
\]
Analogously, one defines $>_{\QQ,\delta}$ on $\QQ$. Thus, $>_{\RR,\delta}$
($>_{\QQ,\delta}$) is well-founded on subsets of $\RR$ ($\QQ$) that are
bounded from below. Therefore, any set $\RR_m$ ($\QQ_m$) could be used
as carrier for polynomial interpretations over $\RR$ ($\QQ$).
However, without loss of generality we may restrict to
$\RR_0$ ($\QQ_0$) because the main argument of
Remark~\ref{rem:isom} also applies to polynomials over $\RR$
($\QQ$), as is already mentioned in \cite{L05}.

\begin{defi}
\label{def:PolyInt_R}
A \emph{polynomial interpretation over $\RR$} for a signature
$\mathcal{F}$ consists of a polynomial $f_\RR \in \RR[\seq{x}]$ for every
$n$-ary function symbol $f \in \mathcal{F}$ and some positive real
number $\delta > 0$ such that
$f_\RR$ is well-defined over $\RR_0$, i.e.,
$f_\RR(\seq{x}) \in \RR_0$ for all $\seq{x} \in \RR_0$.
\end{defi}

Analogously, one defines polynomial interpretations over $\QQ$
by the obvious adaptation of the definition above.
Let $D \in \{ \QQ, \RR \}$.
As for polynomial interpretations over $\NN$,
the pair $\mathcal{D} = (D_0, \{ f_D \}_{f \in \mathcal{F}})$ constitutes
an $\mathcal{F}$-algebra over the carrier $D_0$ due to the
well-definedness of all interpretation functions.
Together with $>_{D_0,\delta}$ and $\geqslant_{D_0}$,
the restrictions of $>_{D,\delta}$ and $\geqslant_D$ to $D_0$,
we obtain an algebra
$(\mathcal{D}, >_{D_0,\delta}, \geqslant_{D_0})$, where
$>_{D_0,\delta}$ is well-founded (on $D_0$) and
${>_{D_0,\delta} \cdot \geqslant_{D_0}} \subseteq {>_{D_0,\delta}}$.
Hence, if for each $f \in \mathcal{F}$, $f_D$ is weakly (strictly)
monotone,
that is, monotone with respect to $\geqslant_{D_0}$ ($>_{D_0,\delta}$),
then $(\mathcal{D}, >_{D_0,\delta}, \geqslant_{D_0})$
is a weakly (strictly) monotone $\mathcal{F}$-algebra.
However, unlike for polynomial interpretations over $\NN$,
strict monotonicity of $(\mathcal{D}, >_{D_0,\delta}, \geqslant_{D_0})$
does not entail weak monotonicity as it can very well be the
case that an interpretation function is monotone with respect to
$>_{D_0,\delta}$ but not with respect to $\geqslant_{D_0}$.

\begin{defi}
Let $D \in \{ \QQ, \RR \}$. A polynomial interpretation over $D$
is said to be \emph{weakly (strictly) monotone} if the algebra
$(\mathcal{D}, >_{D_0,\delta}, \geqslant_{D_0})$ is weakly (strictly)
monotone. Similarly, we say that a polynomial interpretation over $D$ is
\emph{(weakly) compatible} with a TRS $\mathcal{R}$ if
the algebra $(\mathcal{D}, >_{D_0,\delta}, \geqslant_{D_0})$ is (weakly)
compatible with~$\mathcal{R}$.
Finally, a TRS $\mathcal{R}$ is \emph{polynomially terminating over $D$}
if there exists a polynomial interpretation over $D$ that is both
compatible with $\mathcal{R}$ and strictly monotone.
\end{defi}

We conclude this section with a more useful characterization of
monotonicity with respect to the orders $>_{\RR_0,\delta}$ and
$>_{\QQ_0,\delta}$ than the one obtained by specializing
Definition~\ref{def:wma}.
To this end, we note that a function $f\colon \RR_0^n \to \RR_0$ is
strictly monotone in its $i$-th argument with respect to
$>_{\RR_0,\delta}$ if and only if
$f(x_1,\dots,x_i+h,\dots,x_n) - f(x_1,\dots,x_i,\dots,x_n) \geq_\RR \delta$
for all $\seq{x}, h \in \RR_0$ with $h \geq_\RR \delta$.
From this and from the analogous characterization of
$>_{\QQ_0,\delta}$-monotonicity, it is easy to derive the following
lemmata, which will be used in Sections~\ref{sect:nrvsq}
and~\ref{sect:incrementality}.

\begin{lem}
\label{lem:rlinmon}
\label{lem:qlinmon}
\label{lem:qrlinmon}
For $D \in \{ \QQ, \RR \}$ and $\delta \in D_0$ with $\delta > 0$,
the linear polynomial
$f_D(\seq{x}) = a_nx_n + \cdots + a_1x_1 + a_0$
in $D[\seq{x}]$ is monotone in all arguments with respect to
$>_{D_0,\delta}$ and well-defined if and only if $a_0 \geq 0$ and
$a_i \geq 1$ for all $i \in \{ 1, \dots, n \}$.
\qed
\end{lem}

\begin{lem}
\label{lem:qrquadmon}
For $D \in \{ \QQ, \RR \}$ and $\delta \in D_0$ with $\delta > 0$,
the quadratic polynomial $f_D(x) = ax^2 + bx + c$ in $D[x]$ is 
monotone with respect to $>_{D_0,\delta}$ and
well-defined if and only if $a > 0$, $c \geq 0$,
$a\delta + b \geq 1$, and $b \geq 0$ or $4ac - b^2 \geq 0$.
\qed
\end{lem}

In the remainder of this article we will sometimes use the term
``polynomial interpretations with integer coefficients'' as a synonym
for polynomial interpretations over $\NN$. Likewise, the term
``polynomial interpretations with real (rational) coefficients'' refers
to polynomial interpretations over $\RR$ ($\QQ$).

\begin{rem}
\label{Lucas}
Lucas~\cite{L06,L07} considers a different definition of
polynomial termination over $\RR$ ($\QQ$). He allows an
arbitrary subset $A \subseteq \RR$ ($A \subseteq \QQ$) as
interpretation domain,
provided it is bounded from below and unbounded from above.
The definition of well-definedness is modified accordingly.
According to his definition, polynomial termination over $\NN$
trivially implies polynomial interpretations over $\RR$ (and $\QQ$) since
one can take $A = \NN \subseteq \RR$ and $\delta = 1$, in which
case the induced order $>_{A,\delta}$ is the same as the standard
order on $\NN$. Our definitions are based on the understanding
that the interpretation domain together with the underlying order
determine whether one speaks of polynomial interpretations over the
reals, rationals, or integers.
As a consequence, several of the new results obtained in this paper
do not hold in the setting of \cite{L06,L07}.
\end{rem}

\section{Polynomial Termination over the Reals vs.\ the Rationals}
\label{sect:qvsr}

In this section we show that polynomial termination over $\QQ$
implies polynomial termination over $\RR$. The proof is based upon
the fact that polynomials induce continuous functions, whose
behavior at irrational points is completely defined by the values
they take at rational points.

\begin{lem}
\label{lem:continuity}
Let $f\colon \RR^k \to\RR$ be continuous in $\RR^k$.
If $f(\seq[k]{x}) \geq 0$ for all $\seq[k]{x} \in \QQ_0$, then
$f(\seq[k]{x}) \geq 0$ for all $\seq[k]{x} \in \RR_0$.
\end{lem}
\proof
Let $\vec{x} = (\seq[k]{x}) \in \RR_0^k$ and let
$(\vec{x}_n)_{n \in\NN}$ be a sequence of vectors of non-negative
rational numbers $\vec{x}_n \in \QQ_0^k$ whose limit is
$\vec{x}$. Such a sequence exists because $\QQ^k$ is dense in
$\RR^k$. Then
\[
f(\vec{x}) = f(\lim_{n \to \infty} \vec{x}_n) =
\lim_{n \to \infty} f(\vec{x}_n)
\]
by continuity of $f$. Thus, $f(\vec{x})$ is the limit of
$(f(\vec{x}_n))_{n \in \NN}$, which is a sequence of non-negative real
numbers by assumption. Hence, $f(\vec{x})$ is non-negative, too.
\qed

\begin{thm}
\label{thm:qvsr}
If a TRS is polynomially terminating over $\QQ$, then it is
also polynomially terminating over $\RR$.
\end{thm}
\proof
Let $\mathcal{R}$ be a TRS over the signature $\mathcal{F}$ that is
polynomially terminating over $\QQ$.
So there exists some polynomial interpretation $\mathcal{I}$ over $\QQ$
consisting of a positive rational number $\delta$ and a
polynomial $f_\QQ \in \QQ[\seq{x}]$ for every $n$-ary function symbol
$f \in \mathcal{F}$ such that:
\begin{enumerate}
\renewcommand{\theenumi}{\alph{enumi}}
\item
for all $n$-ary $f \in \mathcal{F}$,
$f_\QQ(\seq{x}) \geq 0$ for all $\seq{x} \in \QQ_0$,
\item
for all $f \in \mathcal{F}$,
$f_\QQ$ is strictly monotone with respect to $>_{\QQ_0,\delta}$ in
all arguments,
\item
for every rewrite rule $\ell \to r \in \mathcal{R}$,
$P_\ell >_{\QQ_0,\delta} P_r$ for all
$\seq[m]{x} \in \QQ_0$.
\end{enumerate}
Here $P_\ell$ ($P_r$) denotes the polynomial
associated with $\ell$ ($r$) and the variables
$\seq[m]{x}$ are those occurring in $\ell \to r$.
Next we note that all three conditions are quantified polynomial
inequalities of the shape
``$P(\seq[k]{x}) \geq 0$ for all $\seq[k]{x} \in \QQ_0$'' for
some polynomial $P$ with rational coefficients. This is easy to see
for the first and third condition. As to the second condition,
the function $f_\QQ$ is strictly monotone in its $i$-th argument with
respect to $>_{\QQ_0,\delta}$ if and only if
$f_\QQ(x_1,\dots,x_i+h,\dots,x_n) - f_\QQ(x_1,\dots,x_i,\dots,x_n)
\geq \delta$ for all $\seq{x}, h \in \QQ_0$ with $h \geq \delta$,
which is equivalent to
\[
f_\QQ(x_1,\dots,x_i+\delta+h,\dots,x_n) - f_\QQ(x_1,\dots,x_i,\dots,x_n)
- \delta \geq 0
\]
for all $\seq{x}, h \in \QQ_0$. From
Lemma~\ref{lem:continuity} and the fact that polynomials induce continuous
functions we infer that all these polynomial inequalities do not only
hold in $\QQ_0$ but also in $\RR_0$.
Hence, the polynomial interpretation $\mathcal{I}$
proves termination over $\RR$.
\qed

\begin{rem}\label{rem:qvsr}
Not only does the result established above
show that polynomial termination over $\QQ$ implies polynomial termination
over $\RR$, but it even reveals that the same interpretation applies.
\end{rem}

We conclude this section by showing that
for polynomial interpretations over $\RR$ it suffices to consider real
\emph{algebraic}\footnote{A real number is said to be algebraic if
it is a root of a non-zero polynomial in one variable with rational
coefficients.}
numbers as interpretation domain.
Concerning the use of real algebraic numbers in polynomial
interpretations, in \cite[Section~6]{L07} it is shown that it suffices to
consider polynomials with real algebraic coefficients as interpretations
of function symbols.
Now the obvious question is whether it is also sufficient to consider
only the (non-negative) real algebraic numbers $\RR_{\m{alg}}$ instead of
the entire set $\RR$ of real numbers as interpretation domain. We give an
affirmative answer to this question by extending the result of \cite{L07}.

\begin{thm}
A finite TRS is polynomially terminating over $\RR$ if and only if
it is polynomially terminating over $\RR_{\m{alg}}$.
\end{thm}
\proof
Let $\mathcal{R}$ be a TRS over the signature $\mathcal{F}$ that is
polynomially terminating over $\RR$. 
There exists a positive real number
$\delta$ and a polynomial $f_\RR \in \RR[\seq{x}]$ for every
$n$-ary function symbol $f \in \mathcal{F}$ such that:
\begin{enumerate}
\renewcommand{\theenumi}{\alph{enumi}}
\item
for all $n$-ary $f \in \mathcal{F}$,
$f_\RR(\seq{x}) \geq 0$ for all $\seq{x} \in \RR_0$,
\item
for all $f \in \mathcal{F}$,
$f_\RR$ is strictly monotone with respect to $>_{\RR_0,\delta}$ in
all arguments,
\item
for every rewrite rule $\ell \to r \in \mathcal{R}$,
$P_\ell >_{\RR_0,\delta} P_r$ for all
$\seq[m]{x} \in \RR_0$.
\end{enumerate}
Next we treat $\delta$ as a variable and replace all coefficients of
the polynomials in $\{ f_\RR \mid f \in \mathcal{F} \}$
by distinct variables
$\seq[j]{c}$. Thus, for each $n$-ary function symbol $f \in \mathcal{F}$,
its interpretation function is a parametric polynomial
$f_\RR \in \ZZ[\seq{x},\seq[j]{c}] \subseteq
\ZZ[\seq{x},\seq[j]{c},\delta]$, where all non-zero
coefficients are $1$. As a consequence, we claim that all three
conditions listed above can be expressed as (conjunctions of) quantified
polynomial inequalities of the shape
\begin{equation}
\label{eq:polyconstraints}
p(\seq{x},\seq[j]{c},\delta) \geqslant 0
\quad
\text{for all $\seq{x} \in \RR_0$}
\end{equation}
for some polynomial $p \in \ZZ[\seq{x},\seq[j]{c},\delta]$. This is
easy to see for the first condition.
For the third condition it is a direct consequence of the
nature of the interpretation functions and the usual closure properties
of polynomials. For the second condition we additionally need the fact
that $f_\RR$ is strictly monotone in its $i$-th argument with
respect to $>_{\RR_0,\delta}$ if and only if
$
f_\RR(x_1,\dots,x_i+\delta+h,\dots,x_n) - f_\RR(x_1,\dots,x_i,\dots,x_n)
- \delta \geq 0
$
for all $\seq{x}, h \in \RR_0$. Now any
of the quantified inequalities~\eqref{eq:polyconstraints} can readily
be expressed as a formula in the language of ordered fields with
coefficients in $\ZZ$, where $\seq[j]{c}$ and $\delta$ are the only
free variables. By taking the conjunction of all these formulas,
existentially quantifying $\delta$ and adding the conjunct $\delta > 0$,
we obtain a formula $\Phi$ in the language of ordered fields with free
variables $\seq[j]{c}$ and coefficients in $\ZZ$ (as $\mathcal{R}$
and $\mathcal{F}$ are assumed to be finite). By assumption
there are coefficients
$\seq[j]{C} \in \RR$ such that $\Phi(\seq[j]{C})$ is true in $\RR$, i.e.,
there exists a
satisfying assignment for $\Phi$ in $\RR$ mapping its free
variables $\seq[j]{c}$ to $\seq[j]{C} \in \RR$. In order to prove the
theorem, we first show that there also exists a satisfying assignment
mapping each free variable to a real algebraic number. We
reason as follows. Because real closed fields admit
quantifier elimination (\cite[Theorem~2.77]{BPR06}), there exists a
quantifier-free formula $\Psi$ with free variables $\seq[j]{c}$ and
coefficients in $\ZZ$ that is $\RR$-equivalent to $\Phi$, i.e.,
for all $\seq[j]{y} \in \RR$, $\Phi(\seq[j]{y})$ is true in $\RR$ if
and only if $\Psi(\seq[j]{y})$ is true in $\RR$. Hence, by assumption,
$\Psi(\seq[j]{C})$ is true in $\RR$.
Therefore, the sentence $\exists c_1 \cdots \exists c_j\,\Psi$
is true in $\RR$ as well.
Since both $\RR$ and $\RR_{\m{alg}}$ are real closed fields with
$\RR_{\m{alg}} \subset \RR$ and all coefficients in this sentence
are from $\ZZ \subset \RR_{\m{alg}}$, we may apply the Tarski-Seidenberg
transfer principle (\cite[Theorem~2.80]{BPR06}), from which we infer that
this sentence is true in $\RR$ if
and only if it is true in $\RR_{\m{alg}}$. So there exists an assignment
for $\Psi$ in $\RR_{\m{alg}}$ mapping its free variables $\seq[j]{c}$ to
$\seq[j]{C'} \in \RR_{\m{alg}}$ such that $\Psi(\seq[j]{C'})$ is true
in $\RR_{\m{alg}}$, and hence also in $\RR$ as $\Psi$ is a boolean
combination of atomic formulas in the variables $\seq[j]{c}$ with
coefficients in $\ZZ$. But then $\Phi(\seq[j]{C'})$ is true in $\RR$
as well because of the $\RR$-equivalence of $\Phi$ and $\Psi$.
Another application of the Tarski-Seidenberg transfer principle reveals
that $\Phi(\seq[j]{C'})$ is true in $\RR_{\m{alg}}$,
and therefore the TRS $\mathcal{R}$ is polynomially terminating over
$\RR_{\m{alg}}$ (whose formal definition is the obvious specialization of
Definition~\ref{def:PolyInt_R}). This shows that polynomial termination
over $\RR$ implies polynomial termination over $\RR_{\m{alg}}$.
As the reverse implication can be shown to hold by the same
technique, we conclude that polynomial termination over $\RR$ is
equivalent to polynomial termination over $\RR_{\m{alg}}$.
\qed

\section{Polynomial Termination over the Reals vs.\ the Integers}
\label{sect:nvsr}

As far as the relationship of polynomial interpretations with real,
rational and integer coefficients with regard to termination proving
power is concerned,
Lucas~\cite{L06} managed to prove the following two theorems.%
\footnote{The results of \cite{L06} are actually stronger, cf.\
Remark~\ref{Lucas}.}

\begin{thm}[Lucas, 2006]
\label{thm:qvsn}
There are TRSs that are polynomially terminating over $\QQ$ but not
over $\NN$.
\qed
\end{thm}

\begin{thm}[Lucas, 2006]
\label{thm:rvsq}
There are TRSs that are polynomially terminating over $\RR$ but not
over $\QQ$ or $\NN$.
\qed
\end{thm}

Hence, the extension of the coefficient domain from the integers to the
rational numbers entails the possibility to prove some TRSs
polynomially terminating, which could not be proved polynomially
terminating
otherwise. Moreover, a similar statement holds for the extension of the
coefficient domain from the rational numbers to the real numbers.
Based on these results and the
fact that we have the strict inclusions $\ZZ \subset \QQ \subset \RR$,
it is tempting to believe that polynomial interpretations with real
coefficients properly subsume polynomial interpretations with rational
coefficients, which in turn properly subsume polynomial interpretations
with integer coefficients.
Indeed, the former proposition holds according to
Theorem~\ref{thm:qvsr}.
However, the latter proposition does not hold,
as will be shown in this section.
In particular, we present a TRS that can be proved
terminating by a polynomial interpretation with integer coefficients, but
cannot be proved terminating by a polynomial interpretation
over the reals or rationals.

\subsection{Motivation}

\begin{figure}[htp]
\begin{center}
\begin{tikzpicture}[scale=1.5]
  \draw[->] (-0.2,0) -- (2.4,0) node[right] {$x$};
  \draw[->] (0,-0.1) -- (0,3.7) node[above] {};

  \foreach \y/\ytext in {1/1, 2/2, 3/3, 4/4, 5/5, 6/6, 7/7}
    \draw[yscale=0.5,shift={(0,\y)}] (2pt,0pt) -- (-2pt,0pt) node[left]
    {$\ytext$};

  \foreach \x/\xtext in {0/0, 1/1, 2/2}
    \draw[shift={(\x,0)}] (0pt,2pt) -- (0pt,-2pt) node[below] {$\xtext$};

  \draw[color=red,yscale=0.5] (0,0) parabola bend (0.25,-0.125) (2.1,6.72);

  \draw[color=red,fill=red] (0,0) circle (0.05cm);
  \draw[color=red,fill=red] (1,0.5) circle (0.05cm);
  \draw[color=red,fill=red] (2,3) circle (0.05cm)
    node[below left] {$2x^2-x$ \quad};
\end{tikzpicture}
\end{center}
\caption{The polynomial function $x \mapsto 2x^2 - x$.}
\label{fig:nvsrpoly}
\end{figure}
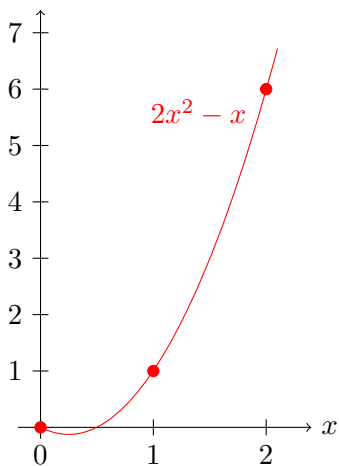

In order to motivate the construction of this particular TRS,
let us first observe that from the viewpoint of number theory there is a
fundamental difference between the integers and the real or rational
numbers. More precisely, the integers are an example of a discrete domain,
whereas both the real and rational numbers are \emph{dense}\footnote{%
Given two distinct real (rational) numbers $a$ and $b$, there exists a
real (rational) number $c$ in between.}
domains. In the context of polynomial interpretations,
the consequences of this major distinction are best explained by an
example. To this end, we consider the polynomial function
$x \mapsto 2x^2 - x$ depicted in
Figure~\ref{fig:nvsrpoly} and assume that we want to use it as
the interpretation of some unary function symbol. Now the point is that
this function is permissible in a polynomial interpretation over $\NN$
as it is both non-negative and strictly monotone over the natural
numbers. However, viewing it
as a function over a
real (rational) variable, we observe that non-negativity is violated
in the open interval $(0,\frac{1}{2})$ (and monotonicity requires a
properly chosen value for $\delta$). Hence, the polynomial function
$x \mapsto 2x^2 - x$ is not permissible in any polynomial interpretation
over $\RR$ ($\QQ$).

Thus, the idea is to design a TRS that enforces an
interpretation of this shape for some unary function symbol, and the tool
that can be used to achieve this is polynomial interpolation.
To this end, let us consider the following scenario, which is
fundamentally based on the assumption that some unary function symbol
$\m{f}$ is interpreted by a quadratic polynomial
$\m{f}(x) = ax^2+bx+c$ with (unknown) coefficients $a$, $b$ and $c$.
Then, by polynomial interpolation, these coefficients are uniquely
determined by the image of $\m{f}$ at three pairwise different
locations; in this way the interpolation constraints
$\m{f}(0) = 0$, $\m{f}(1) = 1$ and $\m{f}(2) = 6$
enforce the interpretation $\m{f}(x) = 2x^2 - x$. Next we encode these
constraints in terms of the TRS $\mathcal{R}$
consisting of the following rewrite rules, where $\m{s}^n(x)$ abbreviates
$
\smash{\underbrace{\m{s}(\m{s}(\cdots\m{s}}_{\text{$n$-times}}(x)\cdots))}
$,
\begin{xalignat*}{2}
\m{s}(\m{0}) &\to \m{f}(\m{0}) \\
\m{s}^2(\m{0}) &\to \m{f}(\m{s}(\m{0})) &
\m{f}(\m{s}(\m{0})) &\to \m{0} \\
\m{s}^7(\m{0}) &\to \m{f}(\m{s}^2(\m{0})) &
\m{f}(\m{s}^2(\m{0})) &\to \m{s}^5(\m{0})
\end{xalignat*}
and consider the following two cases:
polynomial interpretations over $\NN$ on the one hand
and polynomial interpretations over $\RR$ on the other hand.

In the context of polynomial interpretations over $\NN$, we observe that
if we equip the function symbols $\m{s}$ and $\m{0}$ with the (natural)
interpretations $\m{s}_{\NN}(x) = x+1$ and $\m{0}_{\NN} = 0$, then the TRS
$\mathcal{R}$ indeed implements the above interpolation constraints.%
\footnote{In fact, one can even show that $\m{s}_{\NN}(x) = x+1$ is
sufficient for this purpose.}
For example, the constraint
$\m{f}_\NN(1) = 1$ is expressed by $\m{f}(\m{s}(\m{0})) \to \m{0}$ and
$\m{s}^2(\m{0}) \to \m{f}(\m{s}(\m{0}))$. The former encodes
$\m{f}_\NN(1) > 0$,
whereas the latter encodes $\m{f}_\NN(1) < 2$. Moreover, the rule
$\m{s}(\m{0}) \to \m{f}(\m{0})$ encodes $\m{f}_\NN(0) < 1$, which is
equivalent to $\m{f}_\NN(0) = 0$ in the domain of the natural numbers.
Thus, this interpolation constraint can be expressed by a single rewrite
rule, whereas the other two constraints require two rules each. Summing
up, by virtue of the method of polynomial interpolation, we have reduced
the problem of enforcing a specific interpretation for some unary
function symbol to the problem of enforcing natural semantics for the
symbols $\m{s}$ and $\m{0}$.

Next we elaborate on the ramifications of considering the
TRS $\mathcal{R}$ in the context of polynomial interpretations over
$\RR$. To this end, let us assume that the symbols $\m{s}$ and $\m{0}$
are interpreted by $\m{s}_{\RR}(x) = x + s_0$ and $\m{0}_{\RR} = 0$, so
that $\m{s}$ has some
kind of \emph{successor function} semantics. Then the TRS $\mathcal{R}$
translates to the following constraints:
\begin{xalignat*}{2}
s_0 - \delta &\geq_\RR \m{f}_\RR(0) \\
2 s_0 - \delta &\geq_\RR \m{f}_\RR(s_0) &
\m{f}_\RR(s_0) &\geq_\RR 0 + \delta \\
7 s_0 - \delta &\geq_\RR \m{f}_\RR(2 s_0) &
\m{f}_\RR(2 s_0) &\geq_\RR 5 s_0 + \delta
\end{xalignat*}
Hence, $\m{f}_\RR(0)$ is confined to the closed interval
$[0,s_0 - \delta]$, whereas $\m{f}_\RR(s_0)$ is confined to
$[0 + \delta,2 s_0 - \delta]$ and
$\m{f}_\RR(2 s_0)$ to $[5 s_0 + \delta,7 s_0 - \delta]$.
Basically, this means that these constraints do not uniquely determine
the function $\m{f}_\RR$. In other words, the method of polynomial
interpolation does not readily apply to the case of polynomial
interpretations over $\RR$. However, we can make it work. To this end, we
observe that if $s_0 = \delta$, then the above system of inequalities
actually turns into the following system of equations, which can be
viewed as a set of interpolation constraints (parameterized by $s_0$)
that uniquely determine $\m{f}_\RR$:
\begin{xalignat*}{3}
\m{f}_\RR(0) &= 0 &
\m{f}_\RR(s_0) &= s_0 &
\m{f}_\RR(2 s_0) &= 6 s_0
\end{xalignat*}
Clearly, if $s_0 = \delta = 1$, then the symbol $\m{f}$ is fixed to the
interpretation $2x^2 - x$, as was the case in the context of polynomial
interpretations over $\mathbb{N}$ (note that in the latter case
$\delta = 1$ is implicit because of the equivalence
$x >_\NN y \Longleftrightarrow x \geq_\NN y+1$).
Hence, we conclude that once we can manage to design a TRS that enforces
$s_0 = \delta$, we can again leverage the method of polynomial
interpolation to enforce a specific interpretation for some unary
function symbol.
Moreover, we remark that the actual value of $s_0$ is irrelevant
for achieving our goal. That is to say that $s_0$ only serves as a scale
factor in the interpolation constraints determining $\m{f}_\RR$. Clearly,
if $s_0 \neq 1$, then $\m{f}_\RR$ is not fixed to the interpretation
$2x^2 - x$; however, it is still fixed to an interpretation of the same
(desired) shape, as will become clear in the proof of
Lemma~\ref{lem:mainlemma1}.

\subsection{Main Theorem}

In the previous subsection we have presented the basic method that we
use in order to show that polynomial interpretations with real or rational
coefficients do not properly subsume polynomial interpretations with
integer coefficients.
The construction presented there was based on several assumptions, the
essential ones of which are:
\begin{enumerate}
\renewcommand{\theenumi}{\alph{enumi}}
\renewcommand{\labelenumi}{(\theenumi)}
\item \label{Ass1}
The symbol $\m{s}$ had to be interpreted by
a linear polynomial of the shape $x + s_0$.
\item \label{Ass2}
The condition $s_0 = \delta$ was required to hold.
\item \label{Ass3}
The function symbol $\m{f}$ had to be interpreted by a quadratic
polynomial.
\end{enumerate}
Now the point is that one can get rid of all these assumptions
by adding suitable rewrite rules to the TRS $\mathcal{R}$. The
resulting TRS will be referred to as $\mathcal{R}_1$, and it
consists of the rewrite rules given in Table~\ref{tab:trs_r1}.
\begin{table}[tb]
\begin{center}
\fbox{\begin{minipage}[t]{72mm}
\vspace{-1.5ex}
\begin{align}
\m{s}(\m{0}) &\to \m{f}(\m{0}) \label{l1} \\
\m{s}^2(\m{0}) &\to \m{f}(\m{s}(\m{0})) \label{l2} \\
\m{s}^7(\m{0}) &\to \m{f}(\m{s}^2(\m{0})) \label{l3} \\
\m{f}(\m{s}(\m{0}))\vphantom{^2} &\to \m{0} \label{l4} \\
\m{f}(\m{s}^2(\m{0})) &\to \m{s}^5(\m{0}) \label{l5} \\
\m{f}(\m{s}^2(x)) &\to \m{h}(\m{f}(x),\m{g}(\m{h}(x,x))) \label{l6}
\end{align}
\end{minipage}
\begin{minipage}[t]{72mm}
\vspace{-1.5ex}
\begin{align}
\m{f}(\m{g}(x)) &\to \m{g}(\m{g}(\m{f}(x))) \label{r1} \\
\m{g}(\m{s}(x))\vphantom{^2} &\to \m{s}(\m{s}(\m{g}(x))) \label{r2} \\
\m{g}(x)\vphantom{^2} &\to \m{h}(x,x) \label{r3} \\
\m{s}(x)\vphantom{^2} &\to \m{h}(\m{0},x) \label{r4} \\
\m{s}(x)\vphantom{^2} &\to \m{h}(x,\m{0}) \label{r5} \\
\m{h}(\m{f}(x),\m{g}(x))\vphantom{^2} &\to \m{f}(\m{s}(x)) \label{r6}
\end{align}
\end{minipage}}
\end{center}
\caption{The TRS $\mathcal{R}_1$.}
\label{tab:trs_r1}
\end{table}
The rewrite rules \eqref{r1} and \eqref{r2} serve
the purpose of ensuring the first of the above items. Informally,
\eqref{r2} constrains the interpretation of the symbol $\m{s}$ to a
linear polynomial by simple reasoning about the degrees of the left- and
right-hand side polynomials, and \eqref{r1} does the same thing with
respect to $\m{g}$. Because both interpretations are linear, compatibility
with \eqref{r2} can only be achieved if the leading coefficient of
the interpretation of $\m{s}$ is one.

Concerning item~\eqref{Ass3} above, we remark that the tricky part is
to enforce the upper bound of two on the degree of the polynomial
$\m{f}_\RR$
that interprets the symbol $\m{f}$. To this end, we make the
following observation. If
$\m{f}_\RR$
is at most quadratic, then
the function
$\m{f}_\RR(x + s_0) - \m{f}_\RR(x)$
is at most
linear; i.e., there is a linear function
$\m{g}_\RR(x)$
such that
$\m{g}_\RR(x) > \m{f}_\RR(x + s_0) - \m{f}_\RR(x)$,
or equivalently,
$\m{f}_\RR(x) + \m{g}_\RR(x) > \m{f}_\RR(x + s_0)$,
for all values of $x$. This can be encoded in terms of
rule~\eqref{r6}
as soon as the interpretation of $\m{h}$ corresponds to addition of two
numbers. And this is exactly the purpose of rules \eqref{r3},
\eqref{r4} and \eqref{r5}. More precisely, by linearity of the
interpretation of $\m{g}$, we infer from \eqref{r3} that the
interpretation of $\m{h}$ must
have the linear shape $h_2 x + h_1 y + h_0$.
Furthermore, compatibility with \eqref{r4} and \eqref{r5} implies
$h_2 = h_1 = 1$ due to item~\eqref{Ass1} above. Hence, the interpretation
of $\m{h}$ is $x + y + h_0$, and it really models addition of two numbers
(modulo adding a constant).

Next we comment on how to enforce the second of
the above assumptions. To this end, we remark that the hard part is to
enforce the condition $s_0 \leqslant \delta$. The idea is as
follows. First, we consider rule~\eqref{l2}, observing that if
$\m{f}$ is interpreted by a quadratic polynomial $\m{f}_\RR$ and $\m{s}$
by the linear polynomial $x + s_0$, then (the interpretation of) its
right-hand side will eventually become larger than its left-hand side
with growing $s_0$, thus violating compatibility.
In this way, $s_0$ is bounded from above, and the faster the growth of
$\m{f}_\RR$, the lower the bound. The problem with
this statement, however, is that it is only true if $\m{f}_\RR$ is fixed
(which is a priori not the case);
otherwise, for any given value of $s_0$, one can always find a quadratic
polynomial $\m{f}_\RR$
such that compatibility with \eqref{l2} is satisfied.
The parabolic curve associated with $\m{f}_\RR$ only has to be
flat enough. So in order to prevent this, we have to somehow control the
growth of $\m{f}_\RR$. Now that is where rule~\eqref{l6}
comes into play, which basically expresses that if one increases the
argument of $\m{f}_\RR$
by a certain amount (i.e., $2 s_0$), then the value
of the function is guaranteed to increase by a certain minimum amount,
too. Thus, this rule establishes a lower bound on the growth of
$\m{f}_\RR$. And it turns out that if $\m{f}_\RR$
has just the right amount of growth,
then we can readily establish the desired upper bound $\delta$ for $s_0$.

Finally, having presented all the relevant details of our construction,
it remains to formally prove our main claim that the TRS $\mathcal{R}_1$
is polynomially terminating over $\NN$ but not over $\RR$ or $\QQ$.

\begin{lem}
The TRS $\mathcal{R}_1$ is polynomially terminating over $\NN$.
\end{lem}
\proof
We consider the following interpretation:
\[
\m{0}_\NN = 0
\qquad \m{s}_\NN(x) = x + 1
\qquad \m{f}_\NN(x) = 2x^2 - x
\qquad \m{g}_\NN(x) = 4 x + 4
\qquad \m{h}_\NN(x,y) = x + y
\]
Note that the polynomial $2x^2 - x$ is a permissible interpretation
function as it is both non-negative and strictly monotone over the natural
numbers by Lemma~\ref{lem:nquadmon}
(cf.~Figure~\ref{fig:nvsrpoly}). The
rewrite rules of $\mathcal{R}_1$ are compatible with this interpretation
because the resulting inequalities
\begin{xalignat*}{2}
1 &>_\NN 0 &
32x^2 + 60x + 28 &>_\NN 32x^2 - 16x + 20 \\
2 &>_\NN 1 &
4x + 8 &>_\NN 4x + 6 \\
7 &>_\NN 6 &
4x+4 &>_\NN 2x \\
1 &>_\NN 0 &
x+1 &>_\NN x \\
6 &>_\NN 5 &
x+1 &>_\NN x \\
2x^2 + 7x + 6 &>_\NN 2x^2 + 7x + 4 &
2x^2 + 3x + 4 &>_\NN 2x^2 + 3x + 1
\end{xalignat*}
are clearly satisfied for all natural numbers $x$.
\qed

\begin{lem}
\label{lem:mainlemma1}
The TRS $\mathcal{R}_1$ is not polynomially terminating over $\RR$.
\end{lem}
\proof
Let us assume that $\mathcal{R}_1$ is polynomially terminating over $\RR$
and derive a contradiction.
Compatibility with rule \eqref{r2} implies
\[
\deg(\m{g}_\RR(x)) \cdot \deg(\m{s}_\RR(x)) \geq
\deg(\m{s}_\RR(x)) \cdot \deg(\m{s}_\RR(x)) \cdot \deg(\m{g}_\RR(x))
\]
As a consequence, $\deg(\m{s}_\RR(x)) \leq 1$, and because
$\m{s}_\RR$ and $\m{g}_\RR$ must be strictly monotone, we conclude
$\deg(\m{s}_\RR(x)) = 1$. The same reasoning applied to rule \eqref{r1}
yields $\deg(\m{g}_\RR(x)) = 1$. Hence, the symbols $\m{s}$
and $\m{g}$ must be interpreted by linear polynomials. So
$\m{s}_\RR(x) = s_1 x + s_0$ and $\m{g}_\RR(x) = g_1 x + g_0$ with
$s_0, g_0 \in \RR_0$ and, due to Lemma~\ref{lem:rlinmon},
$s_1 \geq_\RR 1$ and $g_1 \geq_\RR 1$.
Then the compatibility constraint imposed by rule \eqref{r2} gives rise
to the inequality
\begin{equation}
\label{eq:comp}
g_1 s_1 x + g_1 s_0 + g_0
>_{\RR_0,\delta}
s_1^2 g_1 x + s_1^2 g_0 + s_1 s_0 + s_0
\end{equation}
which must hold for all non-negative real numbers $x$. This implies
the following condition on the respective leading coefficients:
$g_1 s_1 \geq_\RR s_1^2 g_1$. Because of $s_1 \geq_\RR 1$ and
$g_1 \geq_\RR 1$, this can only hold if $s_1 = 1$. Hence,
$\m{s}_\RR(x) = x + s_0$. This result simplifies \eqref{eq:comp} to
$g_1 s_0 >_{\RR_0,\delta} 2 s_0$, which implies $g_1 s_0 >_\RR 2 s_0$.
From this, we conclude that $s_0 >_\RR 0$ and $g_1 >_\RR 2$.

Now suppose that the function symbol $\m{f}$ were also interpreted by
a linear polynomial $\m{f}_\RR$. Then we could apply the same
reasoning to rule \eqref{r1} because it is structurally
equivalent to \eqref{r2}, thus inferring $g_1 = 1$.
However, this would contradict $g_1 >_\RR 2$;
therefore, $\m{f}_\RR$ cannot be linear.

Next we turn our attention to the rewrite rules
\eqref{r3}, \eqref{r4} and \eqref{r5}.
Because $\m{g}_\RR$ is linear, compatibility with \eqref{r3} constrains
the function $h\colon\RR_0 \to \RR_0, x\mapsto \m{h}_\RR(x,x)$
to be at most linear. This can only be the case if
$\m{h}_\RR$ contains no terms of degree two or higher.
In other words, $\m{h}_\RR(x,y) = h_1 \cdot x + h_2 \cdot y + h_0$,
where $h_0 \in \RR_0$, $h_1 \geq_\RR 1$
and $h_2 \geq_\RR 1$ (cf.\ Lemma~\ref{lem:rlinmon}).
Because of $\m{s}_\RR(x) = x + s_0$, compatibility with
\eqref{r5} implies $h_1 = 1$, and compatibility with
\eqref{r4} implies $h_2 = 1$; thus, $\m{h}_\RR(x,y) = x + y + h_0$.

Using the obtained information in the compatibility constraint associated
with rule \eqref{r6}, we get
\[
\m{g}_\RR(x) + h_0 >_{\RR_0,\delta} \m{f}_\RR(x + s_0) - \m{f}_\RR(x)
\quad \text{for all $x \in \RR_0$.}
\]
This implies that
$\deg(\m{g}_\RR(x) + h_0) \geq \deg(\m{f}_\RR(x + s_0) - \m{f}_\RR(x))$,
which simplifies to
$1 \geq \deg(\m{f}_\RR(x)) - 1$ because $s_0 \neq 0$.
Consequently, $\m{f}_\RR$ must be a quadratic polynomial.
Without loss of generality, let $\m{f}_\RR(x) = ax^2 + bx +c$, subject
to the constraints: $a >_\RR 0$ and $c \geq_\RR 0$ because of
non-negativity (for all $x \in \RR_0$), and $a\delta + b \geq_\RR 1$
because $\m{f}_\RR(\delta) >_{\RR_0,\delta} \m{f}_\RR(0)$
due to strict monotonicity of $\m{f}_\RR$.

Next we consider the compatibility constraint associated with rule
\eqref{l6}, from which we deduce an important auxiliary result.
After unraveling the definitions of
$>_{\RR_0,\delta}$ and the interpretation functions, this constraint
simplifies to
\[
4as_0x + 4a s_0^2 + 2bs_0 \geq_\RR 2g_1 x + g_1 h_0 + g_0 + h_0 + \delta
\quad \text{for all $x \in \RR_0$,}
\]
which implies the following condition on the respective leading
coefficients:
$4 a s_0 \geq_\RR 2 g_1$; from this and $g_1 >_\RR 2$, we conclude
\begin{equation}
\label{eq:as0}
a s_0 >_\RR 1
\end{equation}
and note that $a s_0 = \m{f}'_\RR(\frac{s_0}{2}) - \m{f}'_\RR(0)$.
Hence, $a s_0$
expresses the change of the slopes of the tangents to $\m{f}_\RR$ at the
points $(0,\m{f}_\RR(0))$ and $(\frac{s_0}{2},\m{f}_\RR(\frac{s_0}{2}))$,
and thus
\eqref{eq:as0} actually sets a lower bound on the growth of $\m{f}_\RR$.

Now let us consider the combined compatibility constraint imposed by the
rules \eqref{l2} and \eqref{l4}, namely
$\m{0}_\RR + 2 s_0 >_{\RR_0,\delta} \m{f}_\RR(\m{s}_\RR(\m{0}_\RR))
>_{\RR_0,\delta} \m{0}_\RR$, which implies
$\m{0}_\RR + 2 s_0 \geq_\RR \m{0}_\RR + 2 \delta$
by definition of $>_{\RR_0,\delta}$.
Thus, we conclude $s_0 \geq_\RR \delta$.
In fact, we even have $s_0 = \delta$, which can be derived from
the compatibility constraint of rule \eqref{l2} using the
conditions $s_0 \geq_\RR \delta$,
$a\delta + b \geq_\RR 1$ and
$a s_0 + b \geq_\RR 1$, the combination of the former
two conditions:
\begin{eqnarray*}
\m{0}_\RR + 2 s_0
& >_{\RR_0,\delta} &
\m{f}_\RR(\m{s}_\RR(\m{0}_\RR)) \\
\m{0}_\RR + 2 s_0 -\delta
& \geq_\RR &
\m{f}_\RR(\m{s}_\RR(\m{0}_\RR)) \\
& = &
a(\m{0}_\RR+s_0)^2 + b(\m{0}_\RR + s_0) + c \\
& = &
a\m{0}_\RR^2 + \m{0}_\RR(2a s_0 + b) + a s_0^2 + b s_0 + c \\
& \geq_\RR &
a\m{0}_\RR^2 + \m{0}_\RR + a s_0^2 + b s_0 + c \\
& \geq_\RR &
\m{0}_\RR + a s_0^2 + b s_0 \\
& \geq_\RR & \m{0}_\RR + a s_0^2 + (1 - a\delta)s_0 \\
& = &
\m{0}_\RR + a s_0(s_0 - \delta) + s_0
\end{eqnarray*}
Hence, $\m{0}_\RR + 2 s_0 - \delta \geq_\RR
\m{0}_\RR + a s_0(s_0 - \delta) + s_0$, or equivalently,
$s_0 - \delta \geq_\RR a s_0(s_0 - \delta)$. But because of
\eqref{eq:as0} and $s_0 \geq_\RR \delta$, this inequality can only be
satisfied if:
\begin{equation}
\label{eq:s0=delta}
s_0 = \delta
\end{equation}
This result has immediate consequences concerning the interpretation of
the constant $\m{0}$. To this end, we consider the compatibility
constraint of rule \eqref{r4}, which simplifies to
$s_0 \geq_\RR \m{0}_\RR + h_0 + \delta$.
Because of \eqref{eq:s0=delta} and the fact that $\m{0}_\RR$ and $h_0$
must be non-negative, we conclude $\m{0}_\RR = h_0 = 0$.

Moreover, condition \eqref{eq:s0=delta} is the key to the proof of this
lemma. To this end, we consider the compatibility constraints
associated with the five rewrite rules \eqref{l1}~--~\eqref{l5}:
\begin{xalignat*}{2}
s_0 &>_{\RR_0,s_0} \m{f}_\RR(0) \\
2 s_0 &>_{\RR_0,s_0} \m{f}_\RR(s_0) &
\m{f}_\RR(s_0) &>_{\RR_0,s_0} 0 \\
7 s_0 &>_{\RR_0,s_0} \m{f}_\RR(2 s_0) &
\m{f}_\RR(2 s_0) &>_{\RR_0,s_0} 5 s_0
\end{xalignat*}
By definition of $>_{\RR_0,s_0}$,
these inequalities give rise to the following system of equations:
\begin{xalignat*}{3}
\m{f}_\RR(0) &= 0 &
\m{f}_\RR(s_0) &= s_0 &
\m{f}_\RR(2 s_0) &= 6 s_0
\end{xalignat*}
After unraveling the definition of $\m{f}_\RR$ and substituting
$z := a s_0$, we get a system of linear equations in the unknowns $z$, $b$
and $c$
\begin{xalignat*}{3}
c &= 0 &
z + b &= 1 &
4z + 2b &= 6
\end{xalignat*}
which has the unique solution $z = 2$, $b = -1$ and $c = 0$.
Hence, $\m{f}_\RR$ must have the shape
$\m{f}_\RR(x) = ax^2-x = ax(x - \frac{1}{a})$ in
every compatible polynomial interpretation over $\RR$.
However, this function is not a permissible interpretation for the
function symbol $\m{f}$ because it is not non-negative for all
$x \in \RR_0$. In particular, it is negative in the open interval
$(0,\frac{1}{a})$; e.g. $\m{f}_\RR(\frac{1}{2a}) = -\frac{1}{4a}$.
Hence, $\mathcal{R}_1$
is not compatible with any polynomial interpretation over $\RR$.
\qed

\begin{rem}
In this proof the interpretation of $\m{f}$ is fixed to
$\m{f}_\RR(x) = ax^2-x$, which violates well-definedness in $\RR_0$.
However, this function is obviously well-defined in $\RR_m$ for
a properly chosen negative real number $m$. So what happens if we take this
$\RR_m$ instead of $\RR_0$ as the carrier of a polynomial interpretation?
To this end, we observe that $\m{f}_\RR(0) = 0$ and
$\m{f}_\RR(\delta) = \delta(a \delta - 1) = \delta(a s_0 - 1) = \delta$.
Now let us consider some negative real number $x_0 \in \RR_m$.
We have $\m{f}_\RR(x_0) >_\RR 0$ and thus
$\m{f}_\RR(\delta) - \m{f}_\RR(x_0) <_\RR \delta$,
which means that $\m{f}_\RR$ violates monotonicity with respect to the
order $>_{\RR_m,\delta}$.
\end{rem}

The previous lemma, together with Theorem~\ref{thm:qvsr}, yields the
following corollary.

\begin{cor}
The TRS $\mathcal{R}_1$ is not polynomially terminating over $\QQ$.
\qed
\end{cor}

Finally, combining the material presented in this section,
we establish the following theorem, the main result of this section.

\begin{thm}
\label{thm:nvsr}
There are TRSs that can be proved polynomially terminating over $\NN$,
but cannot be proved polynomially terminating over $\RR$ or $\QQ$.
\qed
\end{thm}

We conclude this section with a remark on the actual choice of the
polynomial serving as the interpretation of the function symbol $\m{f}$.

\begin{rem}
As explained at the beginning of this section,
the TRS $\mathcal{R}_1$ was designed to enforce an interpretation for
$\m{f}$, which is permissible in a polynomial interpretation over $\NN$
but not over $\RR$ ($\QQ$). The interpretation of our choice was the
polynomial $2x^2 - x$.
However, we could have chosen any other polynomial as long as it is
well-defined and strictly monotone over $\NN$ but not over $\RR$ ($\QQ$).
The methods introduced in this section are
general enough to handle any such polynomial. So the actual choice
is not that important.
\end{rem}

\section{Polynomial Termination over the Integers and Reals vs.\ the
Rationals}
\label{sect:nrvsq}

This section is devoted to showing that polynomial termination over $\NN$
and $\RR$ does not imply polynomial termination over $\QQ$.
The proof is constructive, so we give a concrete TRS
having the desired properties.
In order to motivate the construction underlying this particular system,
let us consider the following quantified polynomial inequality
\begin{equation*}\label{eq:motiv}
\tag{$\ast$}
\forall\,x \quad (2x^2 - x) \cdot P(a) \geq 0
\end{equation*}
where $P \in \ZZ[a]$ is a polynomial with integer coefficients, all of
whose roots are irrational and which is positive for some
non-negative integer value of $a$. To be concrete, let us take
$P(a) = a^2 - 2$ and try to satisfy \eqref{eq:motiv} in $\NN$, $\QQ_0$
and $\RR_0$, respectively.
First, we observe that $a := \sqrt{2}$ is a satisfying assignment in
$\RR_0$. Moreover, \eqref{eq:motiv} is also satisfiable in $\NN$ by
assigning $a := 2$, for example, and observing that
the polynomial $2 x^2 - x$ is non-negative for all $x \in \NN$.
However, \eqref{eq:motiv} cannot be satisfied in $\QQ_0$ as
non-negativity of $2 x^2 - x$ does not hold for all $x \in \QQ_0$ and $P$
has no rational roots. To sum up, \eqref{eq:motiv} is
satisfiable in $\NN$ and $\RR_0$ but not in $\QQ_0$. Thus, the basic
idea now is to design a TRS containing some rewrite rule whose
compatibility constraint reduces to a polynomial inequality similar
in nature to \eqref{eq:motiv}.
To this end, we rewrite the inequality
$(2x^2 - x) \cdot (a^2 - 2) \geq 0$ to
\[
2 a^2 x^2 + 2 x \geq 4 x^2 + a^2 x
\]
because now both the left- and right-hand side can be
viewed as a composition of several functions, each of which is strictly
monotone and well-defined. In particular, we identify the following
constituents: $\m{h}(x,y) = x + y$, $\m{r}(x) = 2 x$, $\m{p}(x) = x^2$
and $\m{k}(x) = a^2 x$. Thus, the above inequality can be written in the
form
\begin{equation*}\label{eq:motiv2}
\tag{$\ast\ast$}
\m{h}(\m{r}(\m{k}(\m{p}(x))),\m{r}(x)) \geq
\m{h}(\m{r}(\m{r}(\m{p}(x))),\m{k}(x))
\end{equation*}
which can easily be modeled as a rewrite rule. (Note that
$\m{r}(x)$ is not strictly necessary as $\m{r}(x) = \m{h}(x,x)$, but it
gives rise to a shorter encoding.) And then we also need rewrite rules
that enforce the desired interpretations for the function symbols
$\m{h}$, $\m{r}$, $\m{p}$ and $\m{k}$. For this purpose, we leverage the
techniques presented in the previous section, in particular the method of
polynomial interpolation. The resulting TRS will be referred to as
$\mathcal{R}_2$, and it consists of the rewrite rules
given in Table~\ref{tab:trs_r2}.
\begin{table}[tb]
\begin{center}
\begin{tabular}{@{}c@{\qquad}c@{}}
\begin{minipage}[t]{80mm}
\fbox{\begin{minipage}[t]{80mm}
\vspace{-1.5ex}
\begin{align}
\m{f}(\m{g}(x)) &\to \m{g}^2(\m{f}(x)) \label{rule1} \\
\m{g}(\m{s}(x)) &\to \m{s}^2(\m{g}(x)) \label{rule2} \\
\m{s}(x) &\to \m{h}(\m{0},x) \label{rule4} \\
\m{s}(x) &\to \m{h}(x,\m{0}) \label{rule5} \\
\m{f}(\m{0}) &\to \m{0} \label{rulel1} \\
\m{s}^3(\m{0}) &\to \m{f}(\m{s}(\m{0})) \label{rulel2} \\
\m{f}(\m{s}(\m{0})) &\to \m{s}(\m{0}) \label{rulel4} \\
\makebox[18mm][r]{$\m{h}(\m{f}(x),\m{g}(x))$}
&\to \m{f}(\m{s}(x)) \label{rule6} \\
\m{g}(x) &\to
\makebox[36mm][l]{$\m{h}(\m{h}(\m{h}(\m{h}(x,x),x),x),x)$} \label{rulel7}
\\
\m{f}(\m{s}^2(x)) &\to \m{h}(\m{f}(x),\m{g}(\m{h}(x,x))) \label{rulel6}
\end{align}
\end{minipage}~} \\[1ex]
\fbox{\begin{minipage}[t]{55mm}
\vspace{-1.5ex}
\begin{align}
\makebox[18mm][r]{$\m{s}(\m{0})$} &\to
\m{r}(\m{0}) \label{rr1} \\
\m{s}^3(\m{0}) &\to \m{r}(\m{s}(\m{0})) \label{rr2} \\
\m{r}(\m{s}(\m{0})) &\to \m{s}(\m{0}) \label{rr3} \\
\m{g}(x) &\to \m{r}(x) \label{rr4}
\end{align}
\end{minipage}~}
\end{minipage}
&
\begin{minipage}[t]{60mm}
\fbox{\begin{minipage}[t]{60mm}
\vspace{-1.5ex}
\begin{align}
\m{s}(\m{0}) &\to \m{p}(\m{0}) \label{p1} \\
\m{s}^2(\m{0}) &\to \m{p}(\m{s}(\m{0})) \label{p2} \\
\m{p}(\m{s}(\m{0})) &\to \m{0} \label{p3} \\
\m{s}^5(\m{0}) &\to \m{p}(\m{s}^2(\m{0})) \label{p4} \\
\m{p}(\m{s}^2(\m{0})) &\to \m{s}^3(\m{0}) \label{p5} \\
\makebox[20mm][r]{$\m{h}(\m{p}(x),\m{g}(x))$} &\to
\makebox[20mm][l]{$\m{p}(\m{s}(x))$} \label{p6}
\end{align}
\end{minipage}~} \\[3.3ex]
\fbox{\begin{minipage}[t]{60mm}
\vspace{-1.5ex}
\begin{align}
\m{s}(\m{0}) &\to \m{k}(\m{0}) \label{k1} \\
\makebox[20mm][r]{$\m{s}^2(\m{p}^2(\m{a}))$} &\to
\makebox[20mm][l]{$\m{s}(\m{k}(\m{p}(\m{a})))$} \label{k2} \\
\m{s}(\m{k}(\m{p}(\m{a}))) &\to \m{p}^2(\m{a}) \label{k3} \\
\m{g}(x) &\to \m{k}(x) \label{k4} \\
\m{a} &\to \m{0} \label{a1}
\end{align}
\end{minipage}~} \\[3.3ex]
\hspace*{-25mm}\fbox{\begin{minipage}[t]{85mm}
\vspace{-1.5ex}
\begin{align}
\hspace{-2mm}
\m{s}(\m{h}(\m{r}(\m{k}(\m{p}(x))),\m{r}(x))) &\to
\makebox[25mm][l]{$\m{h}(\m{r}^2(\m{p}(x)),\m{k}(x))$} \label{mainrule}
\end{align}
\end{minipage}~}
\end{minipage} \\
\end{tabular}
\end{center}
\caption{The TRS $\mathcal{R}_2$.}
\label{tab:trs_r2}
\end{table}

Each of the blocks serves a specific purpose. The largest block
consists of the rules \eqref{rule1}~--~\eqref{rulel6}
and is basically a slightly modified version of
the TRS $\mathcal{R}_1$ of Table~\ref{tab:trs_r1}.
These rules ensure that the symbol $\m{s}$ has the semantics of a
successor function $x \mapsto x + s_0$. Moreover, for any
compatible polynomial interpretation
over $\QQ$ ($\RR$), it is guaranteed that $s_0$ is equal to $\delta$,
the minimal step width of the order $>_{\QQ,\delta}$.
In Section~\ref{sect:nvsr},
these conditions were identified as the key requirements
for the method of polynomial interpolation to work in this setting.
Finally, this block also enforces $\m{h}(x,y) = x + y$.
The next block, consisting of the rules \eqref{rr1}~--~\eqref{rr4}, makes
use of polynomial interpolation to achieve $\m{r}(x) = 2 x$. Likewise,
the block consisting of the rules \eqref{p1}~--~\eqref{p6} equips the
symbol $\m{p}$ with the semantics of a squaring function.
And the block \eqref{k1}~--~\eqref{a1} enforces the desired semantics
for the symbol $\m{k}$, i.e., a linear function $x \mapsto k_1 x$
whose slope $k_1$ is proportional to the square of the interpretation of
the constant $\m{a}$.
Finally, the rule \eqref{mainrule} encodes the main idea presented at the
beginning of this section (cf.\ \eqref{eq:motiv2}).

\begin{lem}\label{lem:nrvsq1}
The TRS $\mathcal{R}_2$ is polynomially terminating over $\NN$ and $\RR$.
\end{lem}
\proof
For polynomial termination over $\NN$, the following interpretation
applies:
\begin{gather*}
\m{0}_\NN = 0 \qquad
\m{s}_\NN(x) = x + 1 \qquad
\m{f}_\NN(x) = 3x^2 - 2x + 1 \qquad
\m{g}_\NN(x) = 6x + 6 \\
\m{h}_\NN(x,y) = x + y \qquad
\m{p}_\NN(x) = x^2 \qquad
\m{r}_\NN(x) = 2x \qquad
\m{k}_\NN(x) = 4x \qquad
\m{a}_\NN = 2
\end{gather*}
Note that the polynomial $3x^2 - 2x + 1$ is a permissible
interpretation function by Lemma~\ref{lem:nquadmon}.
Rule \eqref{mainrule} gives rise to the constraint
\[
8x^2 + 2x + 1 >_\NN 4x^2 + 4x
\qquad\iff\qquad
4x^2 - 2x + 1 >_\NN 0
\]
which holds for all $x \in \NN$.
For polynomial termination over $\RR$, we let $\delta = 1$
but we have to modify the interpretation as
$4x^2 - 2x + 1 >_{\RR_0,\delta} 0$ does not hold for all $x \in \RR_0$.
Taking $\m{a}_\RR = \sqrt{2}$, $\m{k}_\RR(x) = 2x$ and
the above interpretations for the other function symbols
establishes polynomial termination over $\RR$. Note that the constraint
$
4x^2 + 2x + 1 >_{\RR_0,\delta} 4x^2 + 2x
$
associated with rule \eqref{mainrule} trivially holds.
Moreover, the functions
$\m{f}_\RR(x) = 3x^2 - 2x + 1$ and $\m{p}_\RR(x) = x^2$ are
strictly monotone with respect to $>_{\RR_0,\delta}$ due to
Lemma~\ref{lem:qrquadmon}.
\qed

\begin{lem}
\label{lem:nrvsq2}
The TRS $\mathcal{R}_2$ is not polynomially terminating over $\QQ$.
\end{lem}
\proof
Let us assume that $\mathcal{R}_2$ is polynomially terminating over $\QQ$
and derive a contradiction. Adapting the reasoning in the proof of
Lemma~\ref{lem:mainlemma1},
we infer from compatibility with the rules
\eqref{rule1}~--~\eqref{rulel7} that $\m{s}_\QQ(x) = x + s_0$,
$\m{g}_\QQ(x) = g_1 x + g_0$, $\m{h}_\QQ(x,y) = x + y + h_0$, and
$\m{f}_\QQ(x) = ax^2 + bx +c$, subject to the following constraints:
\[
s_0 >_\QQ 0 \qquad
g_1 >_\QQ 2 \qquad
g_0, h_0 \in \QQ_0 \qquad
a >_\QQ 0 \qquad
c \geq_\QQ 0 \qquad
a\delta + b \geq_\QQ 1
\]
Next we consider the compatibility constraints associated with the rules
\eqref{rulel7} and \eqref{rulel6}, from which we deduce an important
auxiliary
result. Compatibility with rule \eqref{rulel7} implies the condition
$g_1 \geq_\QQ 5$ on the respective leading coefficients
since $\m{h}_\QQ(x,y) = x + y + h_0$, and compatibility with rule
\eqref{rulel6} simplifies to
\[
4as_0x + 4a s_0^2 + 2bs_0 \geq_\QQ 2g_1 x + g_1 h_0 + g_0 + h_0 + \delta
\quad \text{for all $x \in \QQ_0$,}
\]
from which we infer $4 a s_0 \geq_\QQ 2 g_1$; from this and
$g_1 \geq_\QQ 5$, we conclude $a s_0 >_\QQ 2$.

Now let us consider the combined compatibility constraint imposed by
the rules \eqref{rulel2} and \eqref{rulel4}, namely
$\m{0}_\QQ + 3 s_0 >_{\QQ_0,\delta} \m{f}_\QQ(\m{s}_\QQ(\m{0}_\QQ))
>_{\QQ_0,\delta} \m{0}_\QQ + s_0$, which implies
$\m{0}_\QQ + 3 s_0 \geq_\QQ \m{0}_\QQ + s_0 + 2 \delta$
by definition of $>_{\QQ_0,\delta}$.
Thus, we conclude $s_0 \geq_\QQ \delta$.
In fact, we even have $s_0 = \delta$, which can be derived from
the compatibility constraint of rule \eqref{rulel2} using the
conditions $s_0 \geq_\QQ \delta$,
$a\delta + b \geq_\QQ 1$, $a s_0 + b \geq_\QQ 1$, the combination of the
former
two conditions, and $\m{f}_\QQ(\m{0}_\QQ) \geq_\QQ \m{0}_\QQ + \delta$,
the compatibility constraint of rule \eqref{rulel1}:
\begin{eqnarray*}
\m{0}_\QQ + 3 s_0 -\delta
& \geq_\QQ &
\m{f}_\QQ(\m{s}_\QQ(\m{0}_\QQ))
~=~ \m{f}_\QQ(\m{0}_\QQ) + 2a \m{0}_\QQ s_0  + a s_0^2 + b s_0 \\
& \geq_\QQ &
\m{0}_\QQ + \delta + a s_0^2 + b s_0 \\
& \geq_\QQ &
\m{0}_\QQ + \delta + a s_0^2 + (1 - a \delta) s_0
~=~ \m{0}_\QQ + s_0 + \delta + a s_0 (s_0 - \delta)
\end{eqnarray*}
Hence, $\m{0}_\QQ + 3 s_0 - \delta \geq_\QQ
\m{0}_\QQ + s_0 + \delta + a s_0 (s_0 - \delta)$, or equivalently,
$2 (s_0 - \delta) \geq_\QQ a s_0(s_0 - \delta)$. But since
$a s_0 >_\QQ 2$
and $s_0 \geq_\QQ \delta$, this inequality can only hold if
\begin{equation}
\label{eq:s0=delta1}
s_0 = \delta
\end{equation}
This result has immediate consequences concerning the interpretation of
the constant $\m{0}$. To this end, we consider the compatibility
constraint of rule \eqref{rule4}, which simplifies to
$s_0 \geq_\QQ \m{0}_\QQ + h_0 + \delta$.
Because of \eqref{eq:s0=delta1} and the fact that $\m{0}_\QQ$ and $h_0$
must be non-negative, we conclude $\m{0}_\QQ = h_0 = 0$.

Moreover, as in the proof of Lemma~\ref{lem:mainlemma1},
condition \eqref{eq:s0=delta1}
is the key to the proof of the lemma at hand.
To this end, we consider the compatibility constraints
associated with the rules \eqref{p1}~--~\eqref{p5}.
By definition of $>_{\QQ_0,s_0}$,
these constraints give rise to the following system of equations:
\begin{xalignat*}{3}
\m{p}_\QQ(0) &= 0 &
\m{p}_\QQ(s_0) &= s_0 &
\m{p}_\QQ(2 s_0) &= 4 s_0
\end{xalignat*}
Viewing these equations as polynomial interpolation constraints,
we conclude that no linear polynomial can satisfy them
(because $s_0 \neq 0$). Hence, $\m{p}_\QQ$ must at least be quadratic.
Moreover, by rule \eqref{p6}, $\m{p}_\QQ$ is at most quadratic (using the
same reasoning as for rule \eqref{rule6},
cf.\ the proof of Lemma~\ref{lem:mainlemma1}).
So we let $\m{p}_\QQ(x) = p_2 x^2 + p_1 x + p_0$ in the equations above
and infer the (unique) solution $p_0 = p_1 = 0$ and $p_2 s_0 = 1$,
i.e., $\m{p}_\QQ(x) = p_2 x^2$ with $p_2 \neq 0$.

Next we consider the compatibility constraints associated with the rules
\eqref{rr1}~--~\eqref{rr3}, from which we deduce the interpolation
constraints $\m{r}_\QQ(0) = 0$ and $\m{r}_\QQ(s_0) = 2 s_0$.
Because $\m{g}_\QQ$ is linear, $\m{r}_\QQ$ must be linear, too, for
compatibility with rule \eqref{rr4}. Hence, by polynomial interpolation,
$\m{r}_\QQ(x) = 2 x$.
Likewise, $\m{k}_\QQ$ must be linear for
compatibility with rule \eqref{k4}, i.e., $\m{k}_\QQ(x) = k_1 x + k_0$.
In particular, $k_0 = 0$ due to compatibility with rule \eqref{k1}, and
then the compatibility constraints associated with rule
\eqref{k2} and rule \eqref{k3} yield
$p_2^3 \m{a}_\QQ^4 + 2 s_0 - \delta \geq_\QQ k_1 p_2 \m{a}_\QQ^2 + s_0
\geq_\QQ p_2^3 \m{a}_\QQ^4 + \delta$. But $s_0 = \delta$, hence
$k_1 p_2 \m{a}_\QQ^2 = p_2^3 \m{a}_\QQ^4$, and since $\m{a}_\QQ$ cannot be
zero due to compatibility with rule \eqref{a1}, we obtain
$k_1 = p_2^2 \m{a}_\QQ^2$.
In other words, $\m{k}_\QQ(x) = p_2^2 \m{a}_\QQ^2 x$.

Finally, we consider the compatibility constraint associated with rule
\eqref{mainrule}, which simplifies to
\[
(2 p_2 x^2 - x)((p_2 \m{a}_\QQ)^2 - 2) \geq_\QQ 0 \quad
\text{for all $x \in \QQ_0$.}
\]
However, this inequality is unsatisfiable as the polynomial $2 p_2 x^2 - x$
is negative for some $x \in \QQ_0$ and
$(p_2 \m{a}_\QQ)^2 - 2$ cannot be zero because both $p_2$ and $\m{a}_\QQ$
must be rational numbers.
\qed

Combining the previous two lemmata, we obtain
the main result of this section.

\begin{thm}
\label{thm:nrvsq}
There are TRSs that can be proved polynomially terminating over both $\NN$
and $\RR$, but cannot be proved polynomially terminating over $\QQ$.
\qed
\end{thm}

\section{Incremental Polynomial Termination}
\label{sect:incrementality}

In this section, we consider the possibility of establishing termination
by using polynomial interpretations in an incremental way.
In this setting, which goes back to Lankford~\cite[Example~3]{L79}, one
weakens the compatibility requirement of the interpretation and
the TRS $\mathcal{R}$ under consideration to $P_\ell \geq P_r$ for every
rewrite rule $\ell \to r$ of $\mathcal{R}$ and $P_\ell >_\delta P_r$ for
at least one rewrite rule $\ell \to r$ of $\mathcal{R}$. After removing
those rules of $\mathcal{R}$ satisfying the second condition, one is
free to choose a different interpretation for the remaining rules. This
process is repeated until all rewrite rules have been removed.

\begin{defi}
\label{def:polyint_sn_inc}
For $D \in \{ \NN, \QQ, \RR_{\m{alg}}, \RR \}$ and $n \geqslant 1$,
a TRS $\mathcal{R}$ is said to be
\emph{polynomially terminating over $D$ in $n$ steps} if
either
$n = 1$ and $\mathcal{R}$ is polynomially terminating over $D$ or
$n > 1$ and there exists a polynomial interpretation $\mathcal{P}$ over
$D$ and a non-empty subset $\mathcal{S} \subsetneq \mathcal{R}$ such that
\begin{enumerate}
\item
$\mathcal{P}$ is weakly and strictly monotone,
\item
$\mathcal{R} \subseteq {\geqslant_\mathcal{P}}$ and
$\mathcal{S} \subseteq {>_\mathcal{P}}$, and
\item
\label{def_polyint_sn_3}
$\mathcal{R} \setminus \mathcal{S}$ is polynomially terminating over $D$
in $n-1$ steps.
\end{enumerate}
Furthermore, we call a TRS $\mathcal{R}$
\emph{incrementally polynomially terminating over $D$}
if there exists some
$n \geqslant 1$, such that $\mathcal{R}$ is polynomially
terminating over $D$ in $n$ steps.
\end{defi}

In Section~\ref{subsec:relship_nrvsq_inc} we show that incremental
polynomial termination
over $\NN$ and $\RR$ does not imply incremental polynomial termination
over $\QQ$.
In Section~\ref{subsec:relship_nvsr_inc} we show that incremental
polynomial termination over $\NN$ does not imply incremental polynomial
termination over $\RR$. Below we show that the TRSs
$\mathcal{R}_1$ and $\mathcal{R}_2$ cannot be used for this purpose.
We moreover extend Theorems~\ref{thm:qvsr},
\ref{thm:qvsn}, and \ref{thm:rvsq} to incremental polynomial termination.

\begin{thm}
Let $D \in \{ \NN, \QQ, \RR_\m{alg}, \RR \}$, and let $\mathcal{R}$ be a
TRS. If $\mathcal{R}$ is incrementally polynomially terminating
over $D$, then it is terminating.
\qed
\end{thm}
\proof
Note that the polynomial interpretation $\mathcal{P}$ in
Definition~\ref{def:polyint_sn_inc} is an extended monotone algebra that
establishes relative termination of $\mathcal{S}$ with respect to
$\mathcal{R}$, cf.\ \cite[Theorem~3]{EWZ08}. The result follows by an
easy induction on the number of steps $n$ in
Definition~\ref{def:polyint_sn_inc}.
\qed

For weak monotonicity of univariate quadratic polynomials we use the
following obvious criterion.

\begin{lem}
\label{lem:qrquadwmon}
For $D \in \{ \QQ, \RR \}$,
the quadratic polynomial $f_D(x) = ax^2 + bx + c$ in $D[x]$ is 
weakly monotone if and only if $a >_D 0$ and $b, c \geq_D 0$.
\qed
\end{lem}

We give the full picture of the relationship
between the three notions of incremental polynomial termination
over $\NN$, $\QQ$ and $\RR$, showing that it is essentially
the same as the one depicted in Figure~\ref{fig:summary} for direct
polynomial termination.
However,
we have to replace the TRSs $\mathcal{R}_1$ and $\mathcal{R}_2$ as
the proofs of Lemmata~\ref{lem:mainlemma1} and~\ref{lem:nrvsq2} break
down if we allow incremental termination proofs. In more detail, the
proof of Lemma~\ref{lem:mainlemma1} does not extend because the
TRS $\mathcal{R}_1$ is incrementally polynomially terminating over
$\QQ$.

\begin{lem}
\label{lem:R1_ipt_Q}
The TRS $\mathcal{R}_1$ is incrementally polynomially terminating over
$\QQ$.
\end{lem}
\proof
This can be seen by considering the interpretation
\[
\m{0}_\QQ = 0
\quad \m{s}_\QQ(x) = x + 1
\quad \m{f}_\QQ(x) = x^2 + x
\quad \m{g}_\QQ(x) = 2 x + \tfrac{5}{2}
\quad \m{h}_\QQ(x,y) = x + y
\]
with $\delta = 1$. The rewrite rules of $\mathcal{R}_1$ give rise to
the following inequalities:
\begin{xalignat*}{2}
1 &\geq_\QQ 0 &
4x^2 + 12x + \tfrac{35}{4} &\geq_\QQ 4x^2 + 4x + \tfrac{15}{2} \\
2 &\geq_\QQ 2 &
2x + \tfrac{9}{2} &\geq_\QQ 2x + \tfrac{9}{2} \\
7 &\geq_\QQ 6 &
2x+\tfrac{5}{2} &\geq_\QQ 2x \\
2 &\geq_\QQ 0 &
x+1 &\geq_\QQ x \\
6 &\geq_\QQ 5 &
x+1 &\geq_\QQ x \\
x^2 + 5x + 6 &\geq_\QQ x^2 + 5x + \tfrac{5}{2} &
x^2 + 3x + \tfrac{5}{2} &\geq_\QQ x^2 + 3x + 2
\end{xalignat*}
Removing the rules from $\mathcal{R}_1$ for which the corresponding
constraint remains true after strengthening $\geq_\QQ$ to
$>_{\QQ_0,\delta}$, leaves us with \eqref{l2}, \eqref{r2} and
\eqref{r6}, which are easily handled, e.g.\ by the interpretation
\[
\m{0}_\QQ = 0
\qquad \m{s}_\QQ(x) = x + 1
\qquad \m{f}_\QQ(x) = x
\qquad \m{g}_\QQ(x) = 3 x
\qquad \m{h}_\QQ(x,y) = x + y + 2
\qquad \delta = 1
\]
\qed

Similarly, the TRS $\mathcal{R}_2$ of
Table~\ref{tab:trs_r2}
can be shown to be incrementally polynomially terminating over $\QQ$.
The following result strengthens Theorem~\ref{thm:qvsn}.

\begin{thm}
There are TRSs that are incrementally polynomially terminating over
$\QQ$ but not over $\NN$.
\end{thm}
\proof
Consider the TRS $\mathcal{R}_3$ consisting of the single rewrite
rule
\[
\m{f}(\m{a}) \to \m{f}(\m{g}(\m{a}))
\]
It is easy to see that
$\mathcal{R}_3$ cannot be polynomially terminating over $\NN$.
As the notions of polynomial termination and incremental polynomial
termination coincide for one-rule TRSs, $\mathcal{R}_3$ is not
incrementally polynomially terminating over $\NN$.

The following interpretation establishes polynomial termination over
$\QQ$:
\begin{gather*}
\delta = 1 \qquad
\m{a}_\QQ = \tfrac{1}{2} \qquad
\m{f}_\QQ(x) = 4x    \qquad
\m{g}_\QQ(x) = x^2 
\end{gather*}
To this end, we note that the compatibility constraint associated
with the single rewrite rule gives rise to the inequality
$2 >_{\QQ_0,1} 1$, which holds by definition of $>_{\QQ_0,1}$.
Further note that the interpretation functions are
well-defined and monotone with respect to $>_{\QQ_0,1}$
as a consequence of Lemmata~\ref{lem:qrlinmon} and~\ref{lem:qrquadmon}.
\qed

In fact, the TRS $\mathcal{R}_3$ proves the stronger statement that
there are TRSs which are polynomially terminating over $\QQ$ but not
incrementally polynomially terminating over $\NN$.
Our proof is both
shorter and simpler than the original proof of Theorem~\ref{thm:qvsn}
in \cite[pp.~62--67]{L06}, but see Remark~\ref{Lucas}.

In analogy to Theorem~\ref{thm:qvsr}, incremental polynomial
termination over $\QQ$ implies incremental polynomial termination
over $\RR$.

\begin{thm}
\label{cor:qvsr_inc}
If a TRS is incrementally polynomially terminating over $\QQ$, then
it is also incrementally polynomially terminating over $\RR$.
\end{thm}
\proof
The proof of Theorem~\ref{thm:qvsr}
can be extended with the following statements, which also follow from
Lemma~\ref{lem:continuity}:
\begin{enumerate}
\renewcommand{\theenumi}{\alph{enumi}}
\item
weak monotonicity of $f_\QQ$ with respect to $\geq_\QQ$ implies
weak monotonicity on $\RR_0$ with respect to $\geq_\RR$,
\item
$P_\ell \geq_\QQ P_r$ for all $\seq[m]{x} \in \QQ_0$ implies
$P_\ell \geq_\RR P_r$ for all $\seq[m]{x} \in \RR_0$.
\end{enumerate}
Hence the result follows.
\qed

To show that the converse of Theorem~\ref{cor:qvsr_inc} does not hold,
we consider the TRS $\mathcal{R}_4$
consisting of the rewrite rules of Table~\ref{tab:trs_r4}.

\begin{table}[tb]
\begin{center}
\fbox{
\begin{minipage}[t]{48mm}
\vspace{-1.5ex}
\begin{align}
\m{f}(\m{g}(x)) &\to \m{g}(\m{g}(\m{f}(x))) \tag*{\eqref{r1}} \\
\m{g}(\m{s}(x))\vphantom{^2} &\to \m{s}(\m{s}(\m{g}(x)))
\tag*{\eqref{r2}} \\
\m{g}(x)\vphantom{^2} &\to \m{h}(x,x) \tag*{\eqref{r3}} \\
\m{s}(x)\vphantom{^2} &\to \m{h}(\m{0},x) \tag*{\eqref{r4}}
\end{align}
\end{minipage}
\quad
\begin{minipage}[t]{72mm}
\vspace{-1.5ex}
\begin{align}
\m{s}(x)\vphantom{^2} &\to \m{h}(x,\m{0}) \tag*{\eqref{r5}} \\
\phantom{\m{s}(x)\vphantom{^2}} & \notag \\
\m{k}(\m{k}(\m{k}(x)))\vphantom{^2} &\to \m{h}(\m{k}(x),\m{k}(x))
\label{eq:intrule6} \\
\m{s}(\m{h}(\m{k}(x),\m{k}(x)))\vphantom{^2} &\to \m{k}(\m{k}(\m{k}(x)))
\label{eq:intrule7}
\end{align}
\end{minipage}}
\end{center}
\caption{The TRS $\mathcal{R}_4$.}
\label{tab:trs_r4}
\end{table}

\begin{lem}
\label{lem:trs_R_4_sn}
The TRS $\mathcal{R}_4$ is polynomially terminating over $\RR$.
\end{lem}
\proof
We consider the following interpretation:
\begin{gather*}
\delta = 1
\qquad \m{0}_\RR = 0
\qquad \m{s}_\RR(x) = x + 4
\qquad \m{f}_\RR(x) = x^2 \\
\m{g}_\RR(x) = 3x + 5
\qquad \m{h}_\RR(x,y) = x + y
\qquad \m{k}_\RR(x) = \sqrt{2} x + 1
\end{gather*}
The rewrite rules of $\mathcal{R}_4$ are compatible with this
interpretation because the resulting inequalities
\begin{xalignat*}{2}
9x^2 + 30x + 25 &>_{\RR_0,\delta} 9x^2 + 20 &
x + 4 &>_{\RR_0,\delta} x \\
3x + 17 &>_{\RR_0,\delta} 3x + 13 \\
3x + 5 &>_{\RR_0,\delta} 2x &
2\sqrt{2}x + 3 &>_{\RR_0,\delta} 2\sqrt{2}x + 2 \\
x + 4 &>_{\RR_0,\delta} x &
2\sqrt{2}x + 6 &>_{\RR_0,\delta} 2\sqrt{2}x + 3
\end{xalignat*}
are clearly satisfied for all $x \in \RR_0$.
\qed

It remains to show
that $\mathcal{R}_4$ is not incrementally polynomially terminating over
$\QQ$.
We also show that it is neither incrementally polynomially
terminating over $\NN$. But first we present the following
auxiliary result on a subset of its rules.

\begin{lem}
\label{lem:trs_succadd}
Let $D \in \{ \NN, \QQ, \RR \}$, and let $\mathcal{P}$ be a strictly
monotone
polynomial interpretation over $D$ that is weakly compatible with the
rules \eqref{r1}~--~\eqref{r5}. Then the interpretations of the
symbols $\m{s}$, $\m{h}$ and $\m{g}$ have the shape
\begin{gather*}
\m{s}_D(x) = x + s_0
\qquad \m{h}_D(x,y) = x + y + h_0
\qquad \m{g}_D(x) = g_1 x + g_0
\end{gather*}
where all coefficients are non-negative and $g_1 \geqslant 2$. Moreover,
the interpretation of the symbol $\m{f}$ is at least quadratic.
\end{lem}

\proof
Let the unary symbols $\m{f}$, $\m{g}$ and $\m{s}$ be
interpreted by non-constant polynomials $\m{f}_D(x)$, $\m{g}_D(x)$
and $\m{s}_D(x)$. (Note that strict monotonicity of $\mathcal{P}$
obviously
implies these conditions.) Then the degrees of these polynomials
must be at least $1$, such that weak compatibility with \eqref{r2}
implies
\[
\deg(\m{g}_D(x)) \cdot \deg(\m{s}_D(x)) \geqslant
\deg(\m{s}_D(x)) \cdot \deg(\m{s}_D(x)) \cdot \deg(\m{g}_D(x))
\]
which simplifies to $\deg(\m{s}_D(x)) \leqslant 1$. Hence, we obtain
$\deg(\m{s}_D(x)) = 1$ and, by applying the same reasoning to \eqref{r1},
$\deg(\m{g}_D(x)) = 1$.
So the function symbols $\m{s}$ and $\m{g}$ must be interpreted by
linear polynomials $\m{s}_D(x) = s_1 x + s_0$ and
$\m{g}_D(x) = g_1 x + g_0$, where $s_0, s_1, g_0, g_1 \in D_0$ due to
well-definedness over $D_0$ and $s_1, g_1 > 0$ to make them non-constant.
Then the weak compatibility constraint imposed by \eqref{r2}
gives rise to the inequality
\begin{equation}
\label{eq:comp2}
g_1 s_1 x + g_1 s_0 + g_0
\geqslant_{D_0}
s_1^2 g_1 x + s_1^2 g_0 + s_1 s_0 + s_0
\end{equation}
which must hold for all $x \in D_0$. This implies the following condition
on the respective leading coefficients: $g_1 s_1 \geqslant s_1^2 g_1$.
Due to $s_1, g_1 > 0$, this can only hold if $s_1 \leqslant 1$.
Now suppose that the function symbol $\m{f}$ were also interpreted by a
linear polynomial $\m{f}_D$. Then we could apply the same reasoning to
the rule \eqref{r1} because it is structurally equivalent to \eqref{r2},
thus inferring $g_1 \leqslant 1$. So $\m{f}_D$ cannot
be linear if $g_1 > 1$.

Next we consider the rewrite rules \eqref{r3}, \eqref{r4} and \eqref{r5}.
As $\m{g}_D$ is linear, weak compatibility with \eqref{r3}
implies that the function $\m{h}_D(x,x)$ is at most linear as well.
This can only be the case if the interpretation $\m{h}_D$
is a linear polynomial function $\m{h}_D(x,y) = h_1 x + h_2 y + h_0$,
where $h_0, h_1, h_2 \in D_0$ due to well-definedness over $D_0$. Since
$\m{s}_D(x) = s_1 x + s_0$, weak compatibility with \eqref{r5}
implies $s_1 \geqslant h_1$, and weak compatibility with \eqref{r4}
implies $s_1 \geqslant h_2$. Similarly, we obtain
$g_1 \geqslant h_1 + h_2$ from weak compatibility with \eqref{r3}.

Now if $s_1, h_1, h_2 \geqslant 1$, conditions that are implied by
strict monotonicity of $\m{s}_D$ and $\m{h}_D$
(using Lemma~\ref{lem:qrlinmon}
for $D \in \{ \QQ, \RR \}$),
then we obtain $s_1 = h_1 = h_2 = 1$
and $g_1 \geqslant 2$,
such that
\begin{gather*}
\m{s}_D(x) = x + s_0
\qquad \m{h}_D(x,y) = x + y + h_0
\qquad \m{g}_D(x) = g_1 x + g_0
\end{gather*}
with $g_1 \geqslant 2$, which shows that $\m{f}_D$ cannot be linear.
Due to the fact that all of the above assumptions (on the interpretations
of the symbols $\m{f}$, $\m{g}$, $\m{h}$ and $\m{s}$) follow from strict
monotonicity of $\mathcal{P}$, this concludes the proof.
\qed

With the help of this lemma it is easy to show that the
TRS $\mathcal{R}_4$ is not incrementally polynomially terminating
over $\QQ$ or $\NN$.

\begin{lem}
\label{lem:trs_R_4}
The TRS $\mathcal{R}_4$ is not incrementally polynomially terminating
over $\QQ$ or $\NN$.
\end{lem}

\proof
Let $D \in \{ \NN, \QQ \}$, and let $\mathcal{P}$ be a strictly monotone
polynomial interpretation over $D$ that is weakly compatible
with $\mathcal{R}_4$. Then, by Lemma~\ref{lem:trs_succadd},
the interpretations of the symbols $\m{s}$, $\m{h}$ and $\m{g}$ have
the shape
\begin{gather*}
\m{s}_D(x) = x + s_0
\qquad \m{h}_D(x,y) = x + y + h_0
\qquad \m{g}_D(x) = g_1 x + g_0
\end{gather*}
As the interpretations of the symbols $\m{s}$ and $\m{h}$ are linear,
weak compatibility with \eqref{eq:intrule7} implies that the
interpretation of $\m{k}$ is at most linear as well. Then, letting
$\m{k}_D(x) = k_1 x + k_0$, the weak
compatibility constraints associated with \eqref{eq:intrule6}
and \eqref{eq:intrule7} give rise to the following conditions on the
respective leading coefficients: $2 \geqslant k_1^2 \geqslant 2$.
Hence, $k_1 = \sqrt{2}$, which is not a rational number.
So we conclude that there is no strictly monotone polynomial
interpretation over $\NN$ or $\QQ$ that is weakly compatible with
the TRS $\mathcal{R}_4$. This implies that $\mathcal{R}_4$ is not
incrementally polynomially terminating over $\NN$ or $\QQ$.
\qed

Combining Lemmata~\ref{lem:trs_R_4_sn} and~\ref{lem:trs_R_4},
we obtain the following result.

\begin{cor}
\label{cor:rvsq_inc}
There are TRSs that are incrementally polynomially terminating over
$\RR$ but not over $\QQ$ or $\NN$.
\qed
\end{cor}

As a further consequence of Lemmata~\ref{lem:trs_R_4_sn}
and~\ref{lem:trs_R_4}, we see that the TRS $\mathcal{R}_4$ is
polynomially terminating over $\RR$ but not over $\QQ$ or $\NN$,
which provides an alternative proof of Theorem~\ref{thm:rvsq}.

\subsection{\texorpdfstring{Incremental Polynomial Termination
over $\NN$ and $\RR$ vs.\ $\QQ$}{Incremental Polynomial Termination
over N and R vs.\ Q}}
\label{subsec:relship_nrvsq_inc}

Next we establish the analogon of Theorem~\ref{thm:nrvsq}
in the incremental setting. That is, we show that incremental polynomial
termination over $\NN$ and $\RR$ does not imply incremental
polynomial termination over $\QQ$. Again, we give a concrete TRS
having the desired properties, but unfortunately, as was already
mentioned in the introduction of this section, we cannot reuse the
TRS $\mathcal{R}_2$ directly. Nevertheless, we can and do reuse the
principle idea underlying the construction of $\mathcal{R}_2$
(cf.~\eqref{eq:motiv2}). However, we use a different
method than polynomial interpolation in order to enforce
the desired interpretations for the involved function symbols.
To this end, let us consider the (auxiliary) TRS $\mathcal{S}$
consisting of the
rewrite rules given in Table~\ref{tab:trs_s}.
\begin{table}[tb]
\begin{center}
\fbox{\begin{minipage}[t]{48mm}
\vspace{-1.5ex}
\begin{align}
\m{f}(\m{g}(x)) &\to \m{g}(\m{g}(\m{f}(x))) \tag*{\eqref{r1}} \\
\m{g}(\m{s}(x))\vphantom{^2} &\to \m{s}(\m{s}(\m{g}(x)))
\tag*{\eqref{r2}} \\
\m{g}(x)\vphantom{^2} &\to \m{h}(x,x) \tag*{\eqref{r3}} \\
\m{s}(x)\vphantom{^2} &\to \m{h}(\m{0},x) \tag*{\eqref{r4}} \\
\m{s}(x)\vphantom{^2} &\to \m{h}(x,\m{0}) \tag*{\eqref{r5}}
\end{align}
\end{minipage}}
\quad
\fbox{\begin{minipage}[t]{72mm}
\vspace{-1.5ex}
\begin{align}
\m{k}(x) &\to \m{h}(x,x)
\label{srule1} \\
\m{s}^3(\m{h}(x,x)) &\to \m{k}(x)
\label{srule2} \\
\m{h}(\m{f}(x),\m{k}(x))\vphantom{^3} &\to \m{f}(\m{s}(x))
\label{srule3} \\
\m{f}(\m{s}^2(x))\vphantom{^3} &\to \m{h}(\m{f}(x),\m{k}(\m{h}(x,x)))
\label{srule4} \\
\m{f}(\m{s}(x))\vphantom{^3} &\to \m{h}(\m{f}(x),\m{s}(\m{0}))
\label{srule5} \\
\m{s}^2(\m{0})\vphantom{^3} &\to \m{h}(\m{f}(\m{s}(\m{0})),\m{s}(\m{0}))
\label{srule6}
\end{align}
\end{minipage}}
\end{center}
\caption{The auxiliary TRS $\mathcal{S}$.}
\label{tab:trs_s}
\end{table}
The purpose of this TRS is to equip the symbol $\m{s}$ ($\m{f}$) with
the semantics of a successor (squaring) function and to ensure that
the interpretation of the symbol $\m{h}$ corresponds to the addition
of two numbers. Besides, this TRS will not only be helpful in this
subsection but also in the next one.

\begin{lem}
\label{lem:trs_S}
Let $D \in \{ \NN, \QQ, \RR \}$, and let $\mathcal{P}$ be a strictly
monotone
polynomial interpretation over $D$ that is weakly compatible with the
TRS $\mathcal{S}$. Then
\begin{gather*}
\m{0}_D = 0
\quad \m{s}_D(x) = x + s_0
\quad \m{h}_D(x,y) = x + y \\
\quad \m{g}_D(x) = g_1 x + g_0
\quad \m{k}_D(x) = 2 x + k_0
\quad \m{f}_D(x) = a x^2
\end{gather*}
where $a s_0 = 1$, $g_1 \geqslant 2$ and all coefficients are
non-negative.
\end{lem}

\proof
By Lemma~\ref{lem:trs_succadd}, the interpretations of the
symbols $\m{s}$, $\m{h}$ and $\m{g}$ have the shape
$\m{s}_D(x) = x + s_0$, $\m{h}_D(x,y) = x + y + h_0$ and
$\m{g}_D(x) = g_1 x + g_0$,
where all coefficients are non-negative and $g_1 \geqslant 2$.
Moreover, the interpretation of $\m{f}$ is at least quadratic.

Applying this partial interpretation in \eqref{srule1}
and \eqref{srule2}, we obtain, by weak compatibility, the
inequalities
\[
2 x + h_0 + 3 s_0 \geqslant_{D_0} \m{k}_D(x) \geqslant_{D_0}
2 x + h_0 \quad\text{for all $x \in D_0$,}
\]
which imply $\m{k}_D(x) = 2 x + k_0$ with $k_0 \geqslant 0$ (due
to well-definedness over $D_0$).

Next we consider the rule \eqref{srule4} from which we infer that
$\m{s}_D(x) \neq x$ because otherwise weak compatibility would be
violated; hence, $s_0 > 0$. Then, by weak compatibility
with \eqref{srule3}, we obtain the inequality
\[
\m{k}_D(x) + h_0 \geqslant_{D_0} \m{f}_D(x + s_0) - \m{f}_D(x)
\quad\text{for all $x \in D_0$.}
\]
Now this can only be the case if $\deg(\m{k}_D(x) + h_0) \geqslant
\deg(\m{f}_D(x + s_0) - \m{f}_D(x))$, which simplifies to
$1 \geqslant \deg(\m{f}_D(x)) - 1$ since $s_0 \neq 0$ and $\m{f}_D$
is at least quadratic (hence not constant). 
Consequently, $\m{f}_D$ must be a quadratic polynomial function, that
is, $\m{f}_D(x) = ax^2 + bx +c$ with $a > 0$ (due to well-definedness
over $D_0$). Then the inequalities arising from weak
compatibility with \eqref{srule3} and \eqref{srule4}
simplify to
\begin{align*}
2 x + k_0 + h_0  &\geqslant_{D_0} 2 a s_0 x + a s_0^2 + b s_0 \\
4 a s_0 x + 4 a s_0^2 + 2 b s_0 &\geqslant_{D_0} 4 x + 3 h_0 + k_0
\end{align*}
both of which must hold for all $x \in D_0$. Hence, by looking at
the leading coefficients, we infer that $a s_0 = 1$. Furthermore,
weak compatibility with \eqref{srule5} is satisfied if
and only if the inequality
\[
2 a s_0 x + a s_0^2 + b s_0 \geqslant_{D_0} \m{0}_D + s_0 + h_0
\]
holds for all $x \in D_0$. For $x = 0$, and using the condition
$a s_0 = 1$, we conclude that
$b s_0 \geqslant_{D_0} \m{0}_D + h_0 \geqslant_{D_0} 0$, which
implies that $b \geqslant 0$ as $s_0 > 0$.

Using all the information gathered above, the compatibility constraint
associated with \eqref{srule6} gives rise to the inequality
$0 \geqslant_{D_0} \m{f}_D(\m{0}_D) + 2\, \m{0}_D + b s_0 + h_0$,
all of whose summands on the right-hand side are non-negative as
$b \geqslant 0$ and all interpretation functions must be
well-defined over $D_0$. Consequently, we must have
$\m{0}_D = h_0 = b = c = \m{f}_D(\m{0}_D) = 0$.
\qed

In order to establish the main result of this subsection, we extend the
TRS $\mathcal{S}$ by the rewrite rules given in Table~\ref{tab:trs_r5},
calling the resulting system $\mathcal{R}_5$.
\begin{table}[tb]
\begin{center}
\begin{tabular}{@{~~}c@{\qquad}c@{}}
\begin{minipage}[t]{85mm}
\fbox{\begin{minipage}[t]{80mm}
\vspace{-1.5ex}
\begin{align}
\m{k}(x) &\to \m{r}(x) \label{r5r1} \\
\m{s}(\m{r}(x)) &\to \m{h}(x,x) \label{r5r2} \\
\m{h}(\m{0},\m{0}) &\to \m{r}(\m{0}) \label{r5r3}
\end{align}
\end{minipage}} \\[2.3ex]
\fbox{\begin{minipage}[t]{80mm}
\vspace{-1.5ex}
\begin{align}
\hspace*{-1.5ex} \m{h}(\m{r}(\m{q}(\m{f}(x))),\m{r}(x)) &\to
\m{h}(\m{r}^2(\m{f}(x)),\m{q}(x)) \label{r5mainrule}
\end{align}
\end{minipage}}
\end{minipage}
&
\begin{minipage}[t]{65mm}
\hspace*{-5ex}
\fbox{\begin{minipage}[t]{63mm}
\vspace{-1.5ex}
\begin{align}
\m{g}^2(x) &\to \m{q}(x) \label{r5r4} \\
\m{h}(\m{0},\m{0}) &\to \m{q}(\m{0}) \label{r5r5} \\
\m{f}(\m{f}(\m{m})) &\to
\m{q}(\m{f}(\m{m})) \label{r5r6} \\
\hspace*{-1.5ex} \m{h}(\m{0},\m{q}(\m{f}(\m{m}))) &\to
\m{h}(\m{f}(\m{f}(\m{m})),\m{0}) \label{r5r7} \\
\m{m} &\to \m{s}(\m{0}) \label{r5r8}
\end{align}
\end{minipage}}
\end{minipage}
\\
\end{tabular}
\end{center}
\caption{The TRS $\mathcal{R}_5$ (without the $\mathcal{S}$-rules).}
\label{tab:trs_r5}
\end{table}
As in Section~\ref{sect:nrvsq}, each block serves a
specific purpose. The one made up of \mbox{\eqref{r5r1}~--~\eqref{r5r3}}
enforces the desired semantics
for the symbol $\m{r}$, that is, a linear function $x \mapsto 2 x$
that doubles its input, while the block \eqref{r5r4}~--~\eqref{r5r8}
enforces a linear function $x \mapsto q_1 x$ for the symbol $\m{q}$
whose slope $q_1$ is proportional to the square of the interpretation
of the constant $\m{m}$. Finally, \eqref{r5mainrule} encodes
the main idea of the construction, as mentioned above.

\begin{lem}
\label{lem:trs_R_5_sn}
The TRS $\mathcal{R}_5$ is incrementally polynomially terminating over
$\NN$ and $\RR$.
\end{lem}
\proof
For incremental polynomial termination over $\NN$, we start with the
interpretation
\begin{gather*}
\m{0}_\NN = 0 \quad
\m{s}_\NN(x) = x + 1 \quad
\m{f}_\NN(x) = x^2 \quad
\m{g}_\NN(x) = 3x + 5 \\
\m{h}_\NN(x,y) = x + y \quad
\m{k}_\NN(x) = 2x + 2 \quad
\m{q}_\NN(x) = 4x \quad
\m{r}_\NN(x) = 2x \quad
\m{m}_\NN = 2
\end{gather*}
All interpretation functions are well-defined over $\NN$ and strictly
monotone (i.e.,
monotone with respect to $>_\NN$) as well as weakly monotone (i.e.,
monotone with respect to $\geqslant_\NN$). Moreover, it is easy to
verify that this interpretation is weakly compatible
with $\mathcal{R}_5$. In particular, the rule \eqref{r5mainrule} gives
rise to the constraint
\[
8x^2 + 2x \geqslant_\NN 4x^2 + 4x
\qquad\iff\qquad
2x^2 - x \geqslant_\NN 0
\]
which holds for all $x \in \NN$.
After removing the rules from $\mathcal{R}_5$ for which (strict)
compatibility holds, we are left with the rules \eqref{srule5},
\eqref{srule6}, \eqref{r5r3}, \eqref{r5mainrule} and
\eqref{r5r5}~--~\eqref{r5r7}, all of which can be handled (that
is, removed at once) by the following linear interpretation:
\begin{gather*}
\m{0}_\NN = 0 \quad
\m{s}_\NN(x) = 7x + 2 \quad
\m{h}_\NN(x,y) = x + 2y + 1 \\
\m{f}_\NN(x) = 4x + 2 \quad
\m{q}_\NN(x) = 4x \quad
\m{r}_\NN(x) = x \quad
\m{m}_\NN = 0
\end{gather*}
For incremental polynomial termination over $\RR$, we consider the
interpretation
\begin{gather*}
\delta = 1 \quad
\m{0}_\RR = 0 \quad
\m{s}_\RR(x) = x + 1 \quad
\m{f}_\RR(x) = x^2 \quad
\m{g}_\RR(x) = 3x + 5 \\
\m{h}_\RR(x,y) = x + y \quad
\m{k}_\RR(x) = 2x + 2 \quad
\m{q}_\RR(x) = 2x \quad
\m{r}_\RR(x) = 2x \quad
\m{m}_\RR = \sqrt{2}
\end{gather*}
which is both weakly and strictly monotone according to
Lemmata~\ref{lem:qrquadmon} and~\ref{lem:qrquadwmon}.
So all interpretation functions are
well-defined over $\RR_0$ and monotone with respect to $>_{\RR_0,\delta}$
and $\geqslant_{\RR_0}$. Moreover, one easily verifies that this
interpretation is weakly compatible with $\mathcal{R}_5$. In particular,
the constraint $4x^2 + 2x \geqslant_{\RR_0} 4x^2 + 2x$ associated
with \eqref{r5mainrule} trivially holds. After removing the rules
from $\mathcal{R}_5$ for which (strict) compatibility holds (i.e.,
for which the corresponding constraint remains true after
strengthening $\geqslant_{\RR_0}$ to $>_{\RR_0,\delta}$), we are left
with \eqref{srule5}, \eqref{srule6}, \eqref{r5r3}, \eqref{r5mainrule}
and \eqref{r5r5}~--~\eqref{r5r8}, all of which can be removed at once
by the following linear interpretation:
\begin{gather*}
\delta = 1 \quad
\m{0}_\RR = 0 \quad
\m{s}_\RR(x) = 6x + 2 \quad
\m{f}_\RR(x) = 3x + 2 \\
\m{h}_\RR(x,y) = x + 2y + 1 \quad
\m{q}_\RR(x) = 2x \quad
\m{r}_\RR(x) = x \quad
\m{m}_\RR = 3\rlap{\hbox to 83 pt{\hfill\qEd}}
\end{gather*}

\begin{lem}
\label{lem:trs_R_5}
The TRS $\mathcal{R}_5$ is not incrementally polynomially
terminating over $\QQ$.
\end{lem}
\proof
Let $\mathcal{P}$ be a strictly monotone polynomial interpretation over
$\QQ$ that is weakly compatible with $\mathcal{R}_5$. According to
Lemma \ref{lem:trs_S}, the symbols $\m{0}$,
$\m{s}$, $\m{f}$, $\m{g}$, $\m{h}$ and $\m{k}$ are interpreted as
follows:
\begin{gather*}
\m{0}_\QQ = 0
\quad \m{s}_\QQ(x) = x + s_0
\quad \m{h}_\QQ(x,y) = x + y \\
\quad \m{g}_\QQ(x) = g_1 x + g_0
\quad \m{k}_\QQ(x) = 2 x + k_0
\quad \m{f}_\QQ(x) = a x^2
\end{gather*}
where $s_0, g_1, a > 0$ and $g_0, k_0 \geqslant 0$.

As the interpretation of $\m{k}$ is linear, weak compatibility with the
rule \eqref{r5r1} implies that the interpretation of $\m{r}$ is at most
linear as well, i.e., $\m{r}_\QQ(x) = r_1 x + r_0$ with
$r_0 \geqslant 0$ and $ 2 \geqslant r_1 \geqslant 0$. We also have
$r_1 \geqslant 2$ due to weak compatibility with \eqref{r5r2}
and $0 \geqslant r_0$ due to weak compatibility with \eqref{r5r3};
hence, $\m{r}_\QQ(x) = 2 x$.

Similarly, by linearity of $\m{g}_\QQ$ and weak compatibility
with \eqref{r5r4}, the interpretation of $\m{q}$ must have the
shape $\m{q}_\QQ(x) = q_1 x + q_0$. Then weak compatibility
with \eqref{r5r5} yields $0 \geqslant q_0$; hence,
$\m{q}_\QQ(x) = q_1 x$, $q_1 \geqslant 0$.
Next we note that weak compatibility with \eqref{r5r6}
and \eqref{r5r7} implies that
$\m{f}_\QQ(\m{f}_\QQ(\m{m}_\QQ)) = \m{q}_\QQ(\m{f}_\QQ(\m{m}_\QQ))$,
which evaluates to $a^3 \m{m}_\QQ^4 = a\, q_1 \m{m}_\QQ^2$. From this
we infer
that $q_1 =  a^2 \m{m}_\QQ^2$ as $a > 0$ and $\m{m}_\QQ \geqslant s_0 > 0$
due to weak compatibility with \eqref{r5r8}; i.e.,
$\m{q}_\QQ(x) = a^2 \m{m}_\QQ^2 x$.

Finally, we consider the weak compatibility constraint associated
with \eqref{r5mainrule}, which simplifies to
\[
(2 a x^2 - x)((a\, \m{m}_\QQ)^2 - 2) \geqslant 0 \quad
\text{for all $x \in \QQ_0$.}
\]
However, this inequality is unsatisfiable as the polynomial
$2 a x^2 - x$ is negative for some $x \in \QQ_0$ and
$(a\, \m{m}_\QQ)^2 - 2$ cannot be zero because both $a$
and $\m{m}_\QQ$ must be rational numbers.
So we conclude that there is
no strictly monotone polynomial interpretation over $\QQ$
that is weakly compatible with the TRS $\mathcal{R}_5$.
This implies that $\mathcal{R}_5$ is not incrementally polynomially
terminating over $\QQ$.
\qed

Together, Lemma~\ref{lem:trs_R_5_sn} and Lemma~\ref{lem:trs_R_5}
yield the main result of this subsection.

\begin{cor}
\label{cor:nrvsq_inc}
There are TRSs that are incrementally polynomially terminating over
$\NN$ and $\RR$ but not over $\QQ$.
\qed
\end{cor}

\subsection{\texorpdfstring{Incremental Polynomial Termination
over $\NN$ vs.\ $\RR$}{Incremental Polynomial Termination
over N vs.\ R}}
\label{subsec:relship_nvsr_inc}

In this subsection, we show that there are TRSs that are incrementally
polynomially terminating over $\NN$ but not over $\RR$. For this
purpose, we extend the TRS $\mathcal{S}$ of
Table~\ref{tab:trs_s} by the single rewrite rule
\[
\m{f}(x) \to x
\]
and call the resulting system $\mathcal{R}_6$.

\begin{lem}
\label{lem:trs_R_6_sn}
The TRS $\mathcal{R}_6$ is incrementally polynomially terminating
over $\NN$.
\end{lem}
\proof
First, we consider the interpretation
\begin{gather*}
\m{0}_\NN = 0 \qquad
\m{s}_\NN(x) = x + 1 \qquad
\m{f}_\NN(x) = x^2 \\
\m{h}_\NN(x,y) = x + y \qquad
\m{g}_\NN(x) = 3x + 5 \qquad
\m{k}_\NN(x) = 2x + 2
\end{gather*}
which is both weakly and strictly monotone
as well as weakly compatible
with $\mathcal{R}_6$. In particular, the constraint
$x^2 \geqslant_\NN x$ associated with $\m{f}(x) \to x$
holds for all $x \in \NN$.
Removing the rules from $\mathcal{R}_6$ for which (strict)
compatibility holds leaves us with the rules \eqref{srule5},
\eqref{srule6} and $\m{f}(x) \to x$,
which are easily handled, e.g.~by the linear interpretation
\begin{gather*}
\m{0}_\NN = 0 \qquad
\m{s}_\NN(x) = 3x + 2 \qquad
\m{f}_\NN(x) = 2x + 1 \qquad
\m{h}_\NN(x,y) = x + y\rlap{\hbox to 59 pt{\hfill\qEd}}
\end{gather*}

\begin{lem}
\label{lem:trs_R_6}
The TRS $\mathcal{R}_6$ is not incrementally polynomially terminating over
$\RR$ or $\QQ$.
\end{lem}
\proof
Let $D \in \{ \QQ, \RR \}$, and let $\mathcal{P}$ be a polynomial
interpretation
over $D$ that is weakly compatible with $\mathcal{R}_6$, and in which the
interpretation of the function symbol $\m{f}$ has the shape
$\m{f}_D(x) = a x^2$ with $a > 0$. Then the weak compatibility constraint
$a x^2 \geqslant_{D_0} x$ associated with $\m{f}(x) \to x$
does not hold for all $x \in D_0$ because the polynomial
$a x^2-x = ax\left(x - \tfrac{1}{a}\right)$ is negative in the open
interval $\left(0,\tfrac{1}{a}\right)$.
As the above assumption on the interpretation of $\m{f}$ follows from
Lemma~\ref{lem:trs_S} if $\mathcal{P}$ is strictly monotone, we conclude
that there is no strictly monotone polynomial interpretation over $\RR$
or $\QQ$ that is weakly compatible with the TRS $\mathcal{R}_6$.
This implies that $\mathcal{R}_6$ is not incrementally polynomially
terminating over $\RR$ or $\QQ$.
\qed

Together, Lemma~\ref{lem:trs_R_6_sn} and Lemma~\ref{lem:trs_R_6}
yield the main result of this subsection.

\begin{cor}
\label{cor:nvsr_inc}
There are TRSs that are incrementally polynomially terminating over
$\NN$ but not over $\RR$ or $\QQ$.
\qed
\end{cor}

The results presented in this section can be summarized by
stating that the relationships expressed in Figure~\ref{fig:summary}
remain true for incremental polynomial termination, after
replacing $\mathcal{R}_1$ by $\mathcal{R}_6$ and
$\mathcal{R}_2$ by $\mathcal{R}_5$.

\section{Concluding Remarks}
\label{sect:conclusion}

In this article, we investigated the relationship of polynomial
interpretations with real, rational and integer coefficients with
respect to termination proving power.
In particular, we presented three new results, the first of which
shows that polynomial interpretations
over the reals
subsume polynomial interpretations
over the rationals,
the second of which shows that polynomial interpretations
over the reals or rationals
do not properly subsume polynomial interpretations over the
integers,
a result that comes somewhat unexpected, and the third of which shows
that there are TRSs that can be proved terminating by polynomial
interpretations over the naturals or the reals but not
over the rationals.
These results were extended to incremental termination proofs.
In~\cite{FN12} it is shown how to adapt the results to the dependency
pair framework~\cite{GTSF06,HM07}.

We conclude this article by reviewing
our results in the context of automated termination
analysis, where linear polynomial interpretations, i.e., polynomial
interpretations with all interpretation functions being linear,
play an important role. This naturally raises the question as
to what extent the restriction to linear polynomial interpretations
influences the hierarchy depicted in Figure~\ref{fig:summary},
and in what follows we shall see that it changes considerably.
More precisely, the areas inhabited by the TRSs $\mathcal{R}_1$ and
$\mathcal{R}_2$ become empty, such that polynomial termination by a
linear polynomial interpretation over $\NN$ implies
polynomial termination by a linear polynomial interpretation over $\QQ$,
which in turn implies
polynomial termination by a linear polynomial interpretation over $\RR$.
The latter follows directly from Theorem~\ref{thm:qvsr} and
Remark~\ref{rem:qvsr}, whereas the former is shown below.

\begin{lem}
Polynomial termination by a linear polynomial interpretation over $\NN$
implies polynomial termination by a linear polynomial interpretation
over $\QQ$.
\end{lem}
\proof
Let $\mathcal{R}$ be a TRS that is compatible with a linear
polynomial interpretation $\mathcal{I}$ over $\NN$, where every
$n$-ary function symbol~$\m{f}$ is associated
with a linear polynomial $a_n x_n + \cdots + a_1 x_1 + a_0$.
We show that the same interpretation also establishes polynomial
termination over $\QQ$ with the value of $\delta$ set to one.
To this end, we note that in order to guarantee
strict monotonicity and well-definedness over $\NN$,
the coefficients of the respective interpretation functions
have to satisfy the following conditions:
$a_0 \geqslant 0$ and $a_i \geqslant 1$ for all $i \in \{1,\ldots,n\}$.
Hence, by Lemma~\ref{lem:qlinmon}, we also have well-definedness
over $\QQ_0$ and strict monotonicity with respect to
the order $>_{\QQ_0,1}$.
(Strict monotonicity also follows from \cite[Theorem~2]{L05}.)
Moreover, as $\mathcal{R}$ is compatible with $\mathcal{I}$,
each rewrite rule $\ell \to r \in \mathcal{R}$ satisfies
\begin{equation}
\label{eq:linpoly}
P_\ell - P_r >_\NN 0 \quad \text{for all $\seq[m]{x} \in \NN$,}
\end{equation}
where $P_\ell$ ($P_r$) denotes the polynomial associated with $\ell$ ($r$)
and the variables $\seq[m]{x}$ are those occurring in $\ell \to r$.
Since linear functions are closed under composition, the polynomial
$P_\ell - P_r$ is a linear polynomial $c_m x_m + \cdots + c_1 x_1 + c_0$,
such that \eqref{eq:linpoly} holds if and only if
$c_0 \geqslant 1$ and $c_i \geqslant 0$ for all $i \in \{1,\ldots,m\}$.
However, then we also have
\[
P_\ell - P_r >_{\QQ_0,1} 0 \quad \text{for all $\seq[m]{x} \in \QQ_0$,}
\]
which shows that $\mathcal{R}$ is compatible with the linear polynomial
interpretation $(\mathcal{I},\delta) = (\mathcal{I},1)$ over $\QQ$.
\qed

Hence, linear polynomial interpretations over $\RR$ subsume
linear polynomial interpretations over $\QQ$, which in turn
subsume linear polynomial interpretations over $\NN$, and these
subsumptions are proper due to the results of \cite{L06}, which
were obtained using linear polynomial interpretations.

\section*{Acknowledgements}

We thank Harald Zankl for finding the incremental polynomial
interpretation given in the proof of Lemma~\ref{lem:R1_ipt_Q}.
The comments by the reviewers improved the presentation and
helped to clarify the contributions of Salvador Lucas~\cite{L06}.

 \newcommand{\doi}[1]{\href{http://dx.doi.org/#1}{doi:\nolinkurl{#1}}}
  \newcommand{\noop}[1]{}

\end{document}